\documentclass[11pt,onecolumn,oneside]{osajnl}

\journal{jocn} 

\title{Effects of Eavesdropper on the Performance of Mixed $\eta-\mu$ and DGG Cooperative Relaying System}

\author[1]{Noor Ahmad Sarker}
\author[1]{A. S. M. Badrudduza}
\author[2]{Milton Kumar Kundu}
\author[3]{Imran Shafique Ansari}

\affil[1]{Department of Electronics \& Telecommunication Engineering, Rajshahi University of Engineering \& Technology (RUET), Rajshahi-6204, Bangladesh}
\affil[2]{Department of Electrical \& Computer Engineering, RUET, Rajshahi-6204, Bangladesh}
\affil[3]{James Watt School of Engineering, University of Glasgow, Glasgow G12 8QQ, United Kingdom}

\begin{abstract}
Free-space optical (FSO) channel offers line-of-sight wireless communication with high data rates and high secrecy utilizing unlicensed optical spectrum and also paves the way to the solution of the last-mile access problem. Since atmospheric turbulence is a hindrance to an enhanced secrecy performance, the mixed radio frequency (RF)-FSO system is gaining enormous research interest in recent days. But conventional FSO models except for the double generalized Gamma (DGG) model can not demonstrate secrecy performance for all ranges of turbulence severity.  This reason has led us to propose a dual-hop $\eta-\mu$ and unified DGG mixed RF-FSO network while considering eavesdropping at both RF and FSO hops. The security of these proposed scenarios is investigated in terms of two metrics, i.e., strictly positive secrecy capacity and secure outage probability. Exploiting these expressions, we further investigate how the secrecy performance is affected by various system parameters, i.e., fading, turbulence, and pointing errors. A demonstration is made between heterodyne detection (HD) and intensity modulation/direct detection (IM/DD) techniques while exhibiting superior secrecy performance for HD technique over IM/DD technique. Finally, all analytical results are corroborated via Monte-Carlo simulations.

\quad

\end{abstract}

\begin{document}

\maketitle

\section{Introduction}
\subsection{Background and Related Works}
In recent years, free-space optical (FSO) communication has gained momentous attention in the field of wireless communication. FSO has many advantages over conventional wireless connection techniques due to its high speed, interference immunity, secured configuration, larger bandwidth, etc. It has also been proven to be a cost-effective wireless system by providing a sufficient amount of license-free spectrum to its users. Meanwhile, having a great potentiality of solving spectrum scarcity complications places FSO as a valuable candidate for wireless technologies in the upcoming era.

Several researches has been performed over FSO systems \cite{zhu2002free, haas2003capacity, farid2007outage, belmonte2009capacity, bayaki2009performance, farid2011diversity, ansari2015ergodic, ansari2015performance, safari2008relay, boluda2015ergodic, yang2018asymptotic} to prove its capability for high speed communication. Authors in \cite{zhu2002free} investigated the impact of turbulence-induced fading of an FSO network exploiting spatial diversity techniques with multiple receivers. The expression of capacity-vs-outage-probability was derived in \cite{haas2003capacity} at low and high noise regions where simulation results proved the accuracy of those approximations for a moderate number of apertures. Utilizing slow fading conditions, the FSO channel was investigated in \cite{farid2007outage} based on the effect of natural turbulence and pointing error. Analysis of capacity for optical wireless communication system was again demonstrated in \cite{belmonte2009capacity} using maximal ratio combining (MRC) and selective combining (SC) diversity patterns. Investigation of multiple-input-multiple-output (MIMO) FSO channel was performed in \cite{bayaki2009performance,farid2011diversity}. Performance analysis of FSO system using M\'alaga($\mathcal{M}$) turbulence fading was conducted in \cite{ansari2015ergodic,ansari2015performance}. As FSO is a short-range communication medium, relaying schemes were applied in many research works \cite{safari2008relay, boluda2015ergodic, yang2018asymptotic} to increase its communication range. The authors in \cite{safari2008relay} proposed serial and parallel relaying configuration for both amplify-and-forward (AF) and decode-and-forward (DF) based schemes, and succeeded in mitigating the fading effect. A direct source-to-destination link incorporated with DF relaying system was modeled in \cite{boluda2015ergodic} where ergodic capacity (EC) was analyzed with respect to the fading parameters. The authors in \cite{yang2018asymptotic} proposed a two-way relaying (TWR) scheme undergoing double generalized Gamma (DGG) fading.

As FSO medium is highly unfriendly in conveying information signals over a long distance and in non-line-of-sight (NLoS) conditions, the idea of combined radio frequency-free space optical (RF-FSO) communication is brought forward by many researchers \cite{gupta2018performance, 6952039, vellakudiyan2016channel, vellakudiyan2019performance, 7883900, al2018two, petkovic2018exact, yang2015performance, balti2017aggregate, djordjevic2015mixed, salhab2016power, 7881143, al2019precise, anees2015performance, zedini2014performance, pattanayak2018statistical, palliyembil2018capacity, 7145711, arezumand2017outage, soleimani2015generalized, sharma2016decode, 6966082, yang2017unified, zhang2015unified, 7145973, upadhya2020effect, 10754/134733}. This combined communication technique works in such a way that the RF medium covers up the long-distanced path while the FSO medium fills the remaining short portions of that network. For RF link, Rayleigh fading is a popular model used in several RF-FSO research works \cite{gupta2018performance, vellakudiyan2016channel, vellakudiyan2019performance, al2018two, petkovic2018exact, yang2015performance, balti2017aggregate, djordjevic2015mixed, salhab2016power}. Authors in \cite{gupta2018performance} proposed a DF-based RF-FSO system and analyzed outage probability (OP), EC, bit error rate (BER), and symbol error rate (SER), and performances exhibiting various modulation schemes. A mixed Rayleigh-M\'alaga ($\mathcal{M}$) channel was examined in \cite{vellakudiyan2016channel, vellakudiyan2019performance} and novel expressions for cumulative distribution function (CDF) and channel capacity are derived utilizing fixed and variable gain relaying schemes. A TWR scheme was used in \cite{al2018two} with a multi-user theme. A partial relay selection (PRS) scheme for mixed RF-FSO channel was employed in \cite{petkovic2018exact} to investigate the outage performance of the system. To minimize fading and turbulence effects in mixed RF-FSO system, a variable-gain relaying scheme is proposed in \cite{yang2015performance}. Authors in \cite{balti2017aggregate} introduced a dual-hop RF-FSO model assembled with hardware impairments that created negligible impact in low signal-to-noise ratio (SNR) regime but the larger impact in high SNR regime. A channel state information (CSI) based RF-FSO system with AF relaying was investigated in \cite{djordjevic2015mixed}. The authors in \cite{salhab2016power} investigated mixed Rayleigh-(Gamma-Gamma ($\Gamma\Gamma$)) channel with a multi-relaying scheme considering a multi-user perspective. A lot of interest from the researchers is noticed around considering Nakagami-$m$ fading channel at the RF link of mixed RF-FSO channel \cite{al2019precise, anees2015performance, zedini2014performance, pattanayak2018statistical, palliyembil2018capacity, arezumand2017outage, soleimani2015generalized}. Authors in \cite{al2019precise} proposed a mixed Nakagami-$m$-M\'alaga fading channel and investigated OP, average BER (ABER), and EC at high SNRs based on heterodyne detection (HD) and intensity modulation with direct detection (IM/DD) detection techniques at the receiver. A dual-hop AF-based (Nakagami-$m$)-$\Gamma\Gamma$ fading model was proposed in \cite{anees2015performance,zedini2014performance} to investigate OP, EC, and ABER. A combined (Nakagami-$m$)-DGG system was proposed in \cite{pattanayak2018statistical} where novel expressions for PDF and CDF along with some performance metrics, e.g., OP and ABER were derived. Exact and asymptotic expressions of OP and EC were analyzed in \cite{palliyembil2018capacity}. Ref. \cite{arezumand2017outage} introduced a cognitive cooperative RF-FSO model to analyze the outage performance of the system. A generalized fading pair was proposed in \cite{upadhya2020effect, soleimani2015generalized} showing the effects of atmospheric turbulence, misalignment error, and interference. To gain a better understanding of the mixed RF-FSO link, some researchers introduced generalized fading in RF link \cite{sharma2016decode, yang2017unified, zhang2015unified, upadhya2020effect}. Ref. \cite{sharma2016decode} proposed a mixed $(\eta-\mu)-\Gamma\Gamma$ fading link and analyzed OP, EC, and ABER based on the effects of turbulence and detection types. Identical performance metrics were analyzed in \cite{yang2017unified} while considering ($\eta-\mu$)-M\'alaga combined channel.

Due to the time-varying uncertain nature of the wireless medium, physical layer security (PLS) has always been an important issue. Numerous research has been performed on the secrecy of FSO and mixed RF-FSO networks, as such, \cite{islam2021impact, lopez2015physical, lei2017secrecy, yang2018physical, mandira2021secrecy, abd2017physical, lei2018secrecy, lei2018secrecyoutage, saber2019physical, islam2020secrecy, sarker2020secrecy, lei2020secure, pan2019secrecy}. The effect of correlation along with the pointing error was analyzed in \cite{islam2021impact} considering the presence of eavesdropper at the FSO (scenario I) and RF link (scenario II). The authors in \cite{lopez2015physical} investigated PLS over an FSO link experiencing atmospheric turbulence and analyzed the probability of strictly positive secrecy capacity (SPSC). A mixed RF-FSO system was proposed in \cite{lei2017secrecy} considering both variable and fixed-gain AF-based relaying where the authors derived the expression of average secrecy capacity (ASC) and secure outage probability (SOP) in both exact and asymptotic forms. PLS for mixed ($\eta-\mu$)-M\'alaga channel was studied in \cite{yang2018physical} where authors analyzed average secrecy rate (ASR) and SOP. A more generalized model is studied in \cite{mandira2021secrecy} where the authors analyzed the secrecy performance of the system by deriving the expressions of ASC, SOP, and probability of non-zero secrecy multicast capacity (PNSC). A multi-user-based RF-FSO PLS relaying network was studied in \cite{abd2017physical} where authors analyzed intercept probability (IP) and SOP in both exact and asymptotic forms. The authors in \cite{lei2018secrecy} proposed a secure channel model while considering channel imperfection and analyzed the outage behavior. The SOP was again analyzed in \cite{lei2018secrecyoutage} for a mixed RF-FSO simultaneous wireless information and power transfer (SWIPT) downlink system. Another SWIPT model was proposed in \cite{saber2019physical} for mixed (Nakagami-$m$)-M\'alaga single-input-multiple-output (SIMO) channel where authors analyzed ASC and SOP performance. The authors in \cite{islam2020secrecy} proposed a DF-based generalized Gamma-M\'alaga combined channel with a single eavesdropper at the RF link and analyzed ASC, SOP, and SPSC performances. These performance metrics were again analyzed in \cite{sarker2020secrecy} while considering hyper Gamma(HG)-$\Gamma\Gamma$ AF-based relaying system model. The authors in \cite{lei2020secure} analyzed effective secrecy throughput (EST) for mixed RF-FSO secure system while considering imperfect CSI and transmit antenna selection (TAS) patterns. A DF-based secure Rayleigh-$\Gamma\Gamma$ mixed system was investigated in \cite{pan2019secrecy} while considering an eavesdropper at FSO link, where authors analyzed SPSC and lower bound of SOP.

\subsection{Motivation and Contributions}
Over the years, researchers have proposed a lot of irradiance models for FSO communication. Although the log-normal model is one of the most popular models for its simplicity, it is appropriate for only weak turbulence conditions \cite{alquwaiee2015performance} while facing tractability issues. Recently, the Gamma-Gamma channel model is being utilized for FSO channel modeling quite extensively but a study in \cite{chatzidiamantis2010new} reveals that the double Weibull channel model exhibits more superiority, specifically for moderate and strong turbulence conditions. Hence, for the purpose of unification, in \cite{kashani2015novel}, the author proposed a double generalized Gamma (DGG) model that addresses weak to strong turbulence conditions and eliminates the pitfalls of all other existing FSO channel models. Besides, it is also noted that the PLS analysis while considering generalized fading models in both FSO and RF hops is also infrequent. Moreover, most of the works consider eavesdropper's placement at only the RF or FSO hops and also PLS analysis over the DGG model has not been reported in the literature yet that indicates a large research gap in secrecy analysis over FSO channels applicable in all atmospheric turbulence conditions. Hence, in our proposed RF-FSO system, we consider two generalized fading models namely $\eta-\mu$ and DGG fadings in the RF and FSO hops, respectively. We also consider two eavesdropping scenarios i.e. at the RF and FSO hops. Our main contributions in this work are summarized below:

\begin{enumerate}

\item We first derive the CDF and PDF of the considered dual-hop RF-FSO system via utilizing the CDF and PDF of the SNR of each individual hop. To the best of the authors' knowledge, these expressions possess novelty over other existing systems \cite{sharma2016decode,lei2017secrecy} as both of our proposed fading channels (RF and FSO) convey generic fading characteristics \cite{yacoub2007the,kashani2015novel}.

\item We analyze the secrecy performance of the considered system via deducing expressions of two secrecy performance metrics, i.e., SOP and SPSC applicable in two different eavesdropping scenarios. The analytical outcomes are further validated via Monte-Carlo (MC) simulations. Besides, we also derive the asymptotic SOP expression to acquire more insightful observations on the secure outage performance. Although the authors in \cite{lei2017secrecy, yang2018physical, islam2020secrecy, sarker2020secrecy, abd2017physical, lei2018secrecy, lei2018secrecyoutage, saber2019physical, lei2020secure, pan2019secrecy} proposed similar secure systems, we obtain superiority in our analysis over these researches since the proposed models in this work can analyze the secrecy performance overall turbulence conditions. Also, we demonstrate the results presented in \cite{lei2017secrecy,abd2017physical,pan2019secrecy} can be directly achieved as special cases of our work.

\item We also demonstrate selected analytical outcomes while utilizing system performance metrics showing impacts of the fading parameters of RF channel, atmospheric turbulence, pointing errors, etc. Besides, we also investigate HD and IM/DD techniques to demonstrate the supremacy of the HD technique over the IM/DD technique.

\end{enumerate}

\subsection{Organization}
The remaining parts of this work are arranged as follows. Section II presents the system model and formulation of its CDF expressions. Novel expressions for the SOP and SPSC are derived in Section III. Section IV demonstrates the numerical results of our deduced secrecy metrics. Finally, concluding remarks are provided in Section V.

\section{System Model and Problem Formulation}
\begin{figure}[!h]
\vspace{0mm}
    \centerline{\includegraphics[width=0.8\textwidth,angle =0]{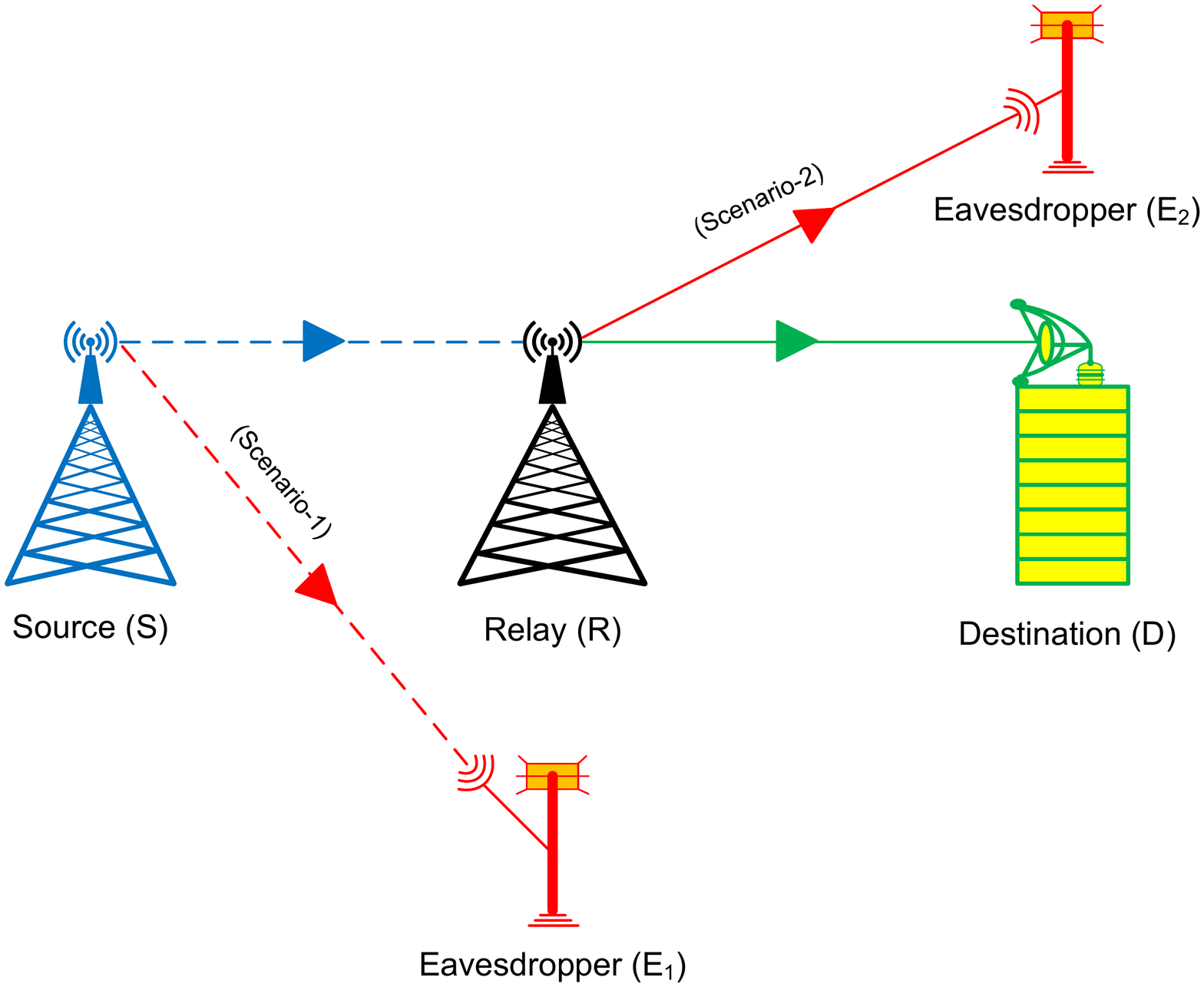}}
    \vspace{0mm }
    \caption{System model incorporating the source (S), relay (R), two eavesdroppers (E$_{1}$ and E$_{2}$), and destination (D).}
    \label{f1}
\end{figure}

We consider a combined RF-FSO system presented in Fig. \ref{f1}, where information is transmitted from source $S$ to destination $D$ through a passable medium relay $R$, where $R$ converts the radio waves into optical waves. We consider DF-based relaying scheme for our proposed system. $S$ consists of one transmitting antenna, $R$ houses one receive antenna and one transmit aperture, and $D$ has an aperture to receive the transmitted optical wave from the relay. The link between $S$ and $R$ is a RF link that is experiencing $\eta-\mu$ fading. On the other hand, $R-D$ link is connected via FSO DGG fading channel. According to the eavesdropper's position, we consider the communication process with experience under the following two scenarios:

\begin{itemize}
\item \textit{\textbf{Scenario-1:}} Eavesdropper $E_{1}$ tries to intercept confidential data transferred through $S-R$ link via another $\eta-\mu$ link. The $R-D$ FSO link is safe from this eavesdropping. Herein, we consider $E_{1}$ consists of one receive antenna.

\item \textit{\textbf{Scenario-2:}} The first hop ($S-R$ link) is totally secure while eavesdropper $E_{2}$ tries to eavesdrop confidential information from $R-D$ link via another DGG fading FSO ($R-E_{2}$) link. Herein, we assume $E_{2}$ houses one receive aperture.
\end{itemize}

\subsection{SNR of Each Link}
For main RF-FSO channel, we represent $\gamma_{r_{0}}$ and $\gamma_{d_{0}}$ as the instantaneous SNRs of $S-R$ and $R-D$ hops, respectively. For eavesdropper channels, $\gamma_{r_{e}}$ and $\gamma_{d_{e}}$ are denoted as the instantaneous SNRs of $S-E_{1}$ link (\textit{Scenario-1}) and $R-E_{2}$ link (\textit{Scenario-2}), respectively. We can mathematically express these SNRs as $\gamma_{r_{0}}=\phi_{s,r}\|\alpha_{s,r}\|^{2}$, $\gamma_{r_{e}}=\phi_{s,e}\|\alpha_{s,e}\|^{2}$, $\gamma_{d_{0}}=\phi_{r,d}\|\alpha_{r,d}\|^{2}$, and $\gamma_{d_{e}}=\phi_{r,e}\|\alpha_{r,e}\|^{2}$, where $\phi_{s,r}$, $\phi_{s,e}$, $\phi_{r,d}$, and $\phi_{r,e}$ denote the average SNRs while $\alpha_{s,r}$, $\alpha_{s,e}$, $\alpha_{r,d}$, and $\alpha_{r,e}$ denote the corresponding channel gains of the $S-R$, $S-E_{1}$, $R-D$, and $R-E_{2}$ links, respectively. For combined $S-R-D$ link, $R$ is utilized for assisting in relaying purpose with the help of channel state information. By considering this fact, SNR of $S-R-D$ and $S-R-E_{2}$ link is defined as \cite[Eq.~(5),]{hasna2004performance}
\begin{subequations}
\begin{align}
&\gamma_{d}=min\left\{\gamma_{r_{0}}, \gamma_{d_{0}}\right\},
\\
&\gamma_{e}=min\left\{\gamma_{r_{0}}, \gamma_{d_{e}}\right\}.
\end{align}
\end{subequations}

\subsection{Secrecy Capacity}

To ensure secure and reliable transmission of information between $S$ and $D$, we need to maintain a rate at which the eavesdropper is unable to wiretap the confidential transmitted data that is known as secrecy rate. To confirm secure transmission for the considered dual-hop communication system in Fig. \ref{f1}, the definition of secrecy capacity (SC) for both scenarios is given below.

\subsubsection{Instantaneous SC for \textit{Scenario-1}}
Considering the first scenario where eavesdropper $E_{1}$ tries to eavesdrop the data from $S$ while experiencing $\eta-\mu$ fading over its link, the instantaneous SC for dual-hop transmission model is defined as \cite[Eq.~(3),]{sarkar2012enhancing}
\begin{align}
\mathcal{T}_{D_{1}}=
    \begin{cases}
      \log_{2}(1+\gamma_{d})-\log_{2}(1+\gamma_{r_{e}}), & \text{if  $\gamma_{d} > \gamma_{r_{e}}$}\\
      0, & \text{if  $\gamma_{d} \leq\gamma_{r_{e}}$}.
    \end{cases}       
\end{align}

\subsubsection{Instantaneous SC for \textit{Scenario-2}}
In the second scenario of Fig. \ref{f1} where the eavesdropper $E_{2}$ tries to eavesdrop the data being transmitted from $R$, two SCs are considered for the two hops (i.e. $S-R$ and $R-D$). As $S-R$ link is not affected by $E_{2}$, instantaneous SC for this link is defined as 
\begin{align} 
\mathcal{T}_{S}=\frac{1}{2}\log_{2}(1+\gamma_{r_{0}}).
\end{align}
For $R-D$ link where data transmission is affected by $R-E_{2}$ eavesdropper link, instantaneous SC is defined as
\begin{align} 
\mathcal{T}_{R}=\biggl[\frac{1}{2}(\log_{2}(1+\gamma_{d_{0}})-\log_{2}(1+\gamma_{d_{e}}))\biggl]^{+},
\end{align}
where $[f]^{+}=max\left\{f,0\right\}$. As \textit{scenario-2} in Fig. \ref{f1} contains DF based relaying, this system is similar to a series configuration where the worst hop acts as the dominating contributor to secrecy capacity of the system. So, the instantaneous SC for the system in Fig. \ref{f1} (\textit{scenario-2}) is defined as \cite[Eq.~(13),]{ai2019physical}
\begin{align}
\label{b1}
\mathcal{T}_{D_{2}}=min(\mathcal{T}_{S},\mathcal{T}_{R}).
\end{align}

\subsection{PDF and CDF of SNR for RF Main Channel}

Considering $\eta-\mu$ fading distribution effecting the main RF channel (i.e. $S-R$ link), the PDF of $\gamma_{r_{o}}$ can be given by \cite[Eq.~(3),]{dacosta2007average}
\begin{align}  
\label{a1}
f_{\gamma_{r_{0}}}(\gamma) =\mathcal{M}_{1}\gamma^{\mu_{0}-\frac{1}{2}}e^{-\mathcal{M
}_{2}\gamma}{I}_{\mu_{0}-\frac{1}{2}}(\mathcal{M}_{3}\gamma),
\end{align}
where $\mathcal{M}_{1}=\frac{2\sqrt{\pi}\mu_{0}^{\mu_{0}+\frac{1}{2}}k_{0}^{\mu_{0}}}{\Gamma(\mu_{0})K_{0}^{\mu_{0}-\frac{1}{2}}\phi_{s,r}^{\mu_{0}+\frac{1}{2}}}$, $\mathcal{M}_{2}=\frac{2k_{0}\mu_{0}}{\phi_{s,r}}$, and $\mathcal{M}_{3}=\frac{2K_{0}\mu_{0}}{\phi_{s,r}}$. Both parameters $k_{0}$ and $K_{0}$ can be explained as $k_{0}=\frac{2+\eta_{0}^{-1}+\eta_{0}}{4}$ and $K_{0}=\frac{\eta_{0}^{-1}-\eta_{0}}{4}$. To satisfy these expressions of $k_{0}$ and $K_{0}$ applicable to main RF channel, the range of $\eta_{0}$ is fixed as $0<\eta_{0}<\infty$ \cite{yacoub2007the, 6692669}. Parameter $\mu_{0}$ represents the fading signal envelope of $S-R$ link with $\mu_{0}>0$ and $\Gamma(.)$ represents Gamma operator \cite[Eq.~(8.310)]{gradshteyn2014table}. $\eta-\mu$ fading distribution has unique generic characteristics that allows it to represent several multipath fading channels as listed in Table \ref{t1}.
\begin{table}[!ht]
\centering
\caption{Special Cases of $\eta-\mu$ Fading Distribution Channel \cite{yacoub2007the}.}
\scalebox{1}{%
\begin{tabular}{|l|c|}
\hline
\multicolumn{1}{|c|}{Channels} & \multicolumn{1}{c|}{$\eta-\mu$ Distribution Parameters}  
\\ 
\hline
Hoyt / Nakagami-$q$  & $\eta=q^{2}, \mu=0.5$    \\ 
One-sided Gaussian & $\eta\rightarrow0$ (or $\eta\rightarrow\infty$), $\mu=0.5$        \\ 
Rayleigh & $\eta\rightarrow0$ (or $\eta\rightarrow\infty$), $\mu=1$        \\ 
Nakagami-$m$ & $\eta\rightarrow0$ (or $\eta\rightarrow\infty$), $\mu=m$
 \\ 
 \hline
\end{tabular}}
\label{t1}
\end{table}
Since the value of $\mu$ is considered as integer in most of the works \cite{yang2018physical}, \eqref{a1} is rewritten as
\begin{align}  
\label{a2}
f_{\gamma_{r_{0}}}(\gamma)=\mathcal{A}\sum_{N_{0}=1}^{2}\sum_{v=0}^{\mu_{0}-1}X_{N_{0},v}\gamma^{\mu_{0}-v-1}e^{-l_{N_{0}}\gamma},
\end{align}
where $\mathcal{A}=\frac{k_{0}^{\mu_{0}}}{K_{0}^{\mu_{0}}\Gamma(\mu_{0})}$, $X_{1,v}=\frac{\Gamma(\mu_{0}+v)(\mu_{0})^{\mu_{0}-v}}{v!\Gamma(\mu_{0}-v)4^{v}\phi_{s,r}^{\mu_{0}-v}K_{0}^{v}}(-1)^{v}$, $X_{2,v}=\frac{\Gamma(\mu_{0}+v)(\mu_{0})^{\mu_{0}-v}}{v!\Gamma(\mu_{0}-v)4^{v}\phi_{s,r}^{\mu_{0}-v}K_{0}^{v}}(-1)^{\mu_{0}}$, $l_{1}=\frac{2\mu_{0}(k_{0}-K_{0})}{\phi_{s,r}}$, and $l_{2}=\frac{2\mu_{0}(k_{0}+K_{0})}{\phi_{s,r}}$. The CDF of $\gamma_{r_{0}}$ can be expressed as \cite[Eq.~(4),] {yang2018physical}
\begin{align}  
\label{a3}
F_{\gamma_{r_{0}}}(\gamma)=1-\mathcal{A}\sum_{N_{0}=1}^{2}\sum_{v=0}^{\mu_{0}-1}\sum_{x=0}^{\mu_{0}-v-1}\frac{\gamma^{x}}{x!}e^{-l_{N_{0}}\gamma}l_{N_{0}}^{x}Y_{N_{0},v},
\end{align}
where $Y_{1,v}=\frac{\Gamma(\mu_{0}+v)(-1)^{v}K_{0}^{-v}}{v!2^{\mu_{0}+v}(k_{0}-K_{0})^{\mu_{0}-v}}$, and $Y_{2,v}=\frac{\Gamma(\mu_{0}+v)(-1)^{\mu_{0}}K_{0}^{-v}}{v!2^{\mu_{0}+v}(k_{0}+K_{0})^{\mu_{0}-v}}$.
\subsection{PDF and CDF of SNR for FSO Main Channel}

Considering DGG fading experience over our proposed FSO hop, the PDF of $R-D$ link is defined as \cite[Eq.~(12),]{alquwaiee2015performance}
\begin{align}  
\label{a4}
f_{\gamma_{d_{0}}}(\gamma)=\frac{\mathcal{B}_{1}}{s_{0}\gamma}G_{1,\lambda_{1}+\lambda_{2}+1}^{\lambda_{1}+\lambda_{2}+1,0}\biggl[\mathcal{B}_{2}t^{\tau}\biggl(\frac{\gamma}{U_{d}}\biggl)^{\frac{\tau}{s_{0}}}\biggl |
 \begin{array}{c}
 j_{2}\\
 j_{1}\\
\end{array}\biggl],
\end{align}
where $\mathcal{B}_{1}=\frac{\epsilon^{2}\lambda_{2}^{b_{1}-\frac{1}{2}}\lambda_{1}^{b_{2}-\frac{1}{2}}(2\pi)^{1-\frac{\lambda_{1}+\lambda_{2}}{2}}}{\Gamma(b_{1})\Gamma(b_{2})}$, $\mathcal{B}_{2}=\frac{b_{1}^{\lambda_{2}}b_{2}^{\lambda_{1}}}{\lambda_{1}^{\lambda_{1}}\lambda_{2}^{\lambda_{2}}\Omega_{1}^{\lambda_{2}}\Omega_{2}^{\lambda_{1}}}$, $t=\frac{\mathcal{B}_{1}\zeta}{(1+\epsilon^{2})\mathcal{B}_{2}^{\frac{1}{\alpha_{2}\lambda_{1}}}}$, and $\tau=a_{2}\lambda_{1}$. The values of $a_{1}$, $a_{2}$, $\Omega_{1}$, and $\Omega_{2}$ are mathematically calculated that are identified by the variances related to large-scale and small-scale fluctuations \cite{al2001mathematical}. The two shaping parameters $b_{1}$ and $b_{2}$ define fading characteristics induced via turbulence conditions \cite{kashani2015novel}. Two positive integers $\lambda_{1}$ and $\lambda_{2}$ are defined such that $\frac{\lambda_{1}}{\lambda_{2}}=\frac{a_{1}}{a_{2}}$ \cite{alquwaiee2015performance}, $s_{0}$ represents two types of detections utilized at $D$ for receiving optical signals ($s_{0}=1$ symbolizes HD technique and $s_{0}=2$ represents IM/DD technique), and $\epsilon$ acts as the indicator for pointing error in the FSO channel that is actually a ratio between the width of equivalent signal beam and the jitter of pointing error displacement \cite{ansari2015performance}. The electrical SNR for DGG fading model over $R-D$ link is defined as $U_{d}=\frac{\left\{\Lambda_{0}E[I_{0}]\right\}^{s_{0}}}{P_{0}}$, respectively. Here, $\Lambda_{0}$, $I_{0}$, and $P_{0}$ denote photoelectric conversion coefficient, receiver irradiance, and number of sample apertures, respectively \cite{tsiftsis2009optical}. Hence, the trivial relationship between $\phi_{r,d}$ and $U_{d}$ can be addressed as $\frac{E[I_{0}^{2}]}{E[I_{0}]^{2}}\triangleq\iota_{0}^{2}+1$, where $\iota_{0}^{2}$ represents the scintillation index \cite{niu2013error}. Parameter $\zeta$ is expressed as $\zeta=\prod_{i=1}^{\lambda_{1}+\lambda_{2}}\Gamma(\frac{1}{a_{2}\lambda_{1}}+\Psi_{i})$, where $\Psi_{q}$ denotes the $q$-th term of $\Psi$ \cite{alquwaiee2015performance}. The terms $\Psi$, $j_{1}$, and $j_{2}$ are expressed as
\begin{align}  
\nonumber
\Psi&=\Delta(\lambda_{2}:b_{1}), \Delta(\lambda_{1}:b_{2}),
\\ 
\nonumber
j_{1}&=\frac{\epsilon^{2}}{a_{2}\lambda_{1}}, \Delta(\lambda_{2}:b_{1}), \Delta(\lambda_{1}:b_{2}),
\\\nonumber
\nonumber
j_{2}&=\frac{a_{2}\lambda_{1}+\epsilon^{2}}{a_{2}\lambda_{1}},
\end{align}
and the symbol notation $\Delta(p:q)$ including $p$ terms is defined as
\begin{align}
\nonumber
\Delta(p:q)=\frac{q}{p},\frac{q+1}{p},\cdots,\frac{q+p-1}{p}.
\end{align}
The DGG turbulence model is a generic fading model for FSO communications thereby housing several classical fading models within itself as listed in Table \ref{t2}.
\begin{table}[!h]
\centering
\caption{Special Cases of DGG Turbulence Fading Channel \cite{kashani2015novel}.}
\scalebox{1}{%
\begin{tabular}{|l|c|}
\hline
\multicolumn{1}{|c|}{Channels} & \multicolumn{1}{c|}{DGG turbulence parameters}  
\\ 
\hline
Double-Weibull & $b_{1}=b_{2}=1$    \\ 
$\Gamma\Gamma$ & $a_{1}=a_{2}=\Omega_{1}=\Omega_{2}=1$        \\ 
Lognormal & $a_{1}\rightarrow0, a_{2}\rightarrow0, b_{1}\rightarrow\infty, b_{2}\rightarrow\infty$        \\ 
$K$ distribution & $a_{1}=a_{2}=b_{2}=\Omega_{1}=\Omega_{2}=1$
 \\ 
 \hline
\end{tabular}}
\label{t2}
\end{table}
Hence, this is one of the most popular fading model that has attracted many OWC researchers. Moreover, the CDF of $\gamma_{d_{0}}$ is defined as \cite[Eq.~(14),]{alquwaiee2015performance}
\begin{align}
\label{a5}
F_{\gamma_{d_{0}}}(\gamma)=\mathcal{B}_{3}G_{s_{0}+1,\delta_{0}+1}^{\delta_{0},1}\biggl[\mathcal{B}_{4}\biggl(\frac{\gamma}{U_{d}}\biggl)^{\tau}\biggl |
 \begin{array}{c}
 1, j_{3}\\
 j_{4}, 0\\
\end{array}\biggl],
\end{align}
where $\mathcal{B}_{3}=\frac{\epsilon^{2}\lambda_{2}^{b_{1}-\frac{1}{2}}\lambda_{1}^{b_{2}-\frac{1}{2}}(2\pi)^{1-\frac{s_{0}(\lambda_{1}+\lambda_{2})}{2}}s_{0}^{b_{1}+b_{2}-2}}{a_{2}\lambda_{1}\Gamma(b_{1})\Gamma(b_{2})}$, $B_{4}=\left(\frac{\mathcal{B}_{2}t^{a_{2}\lambda_{1}}}{s_{0}^{\lambda_{1}+\lambda_{2}}}\right)^{s_{0}}$, and $\delta_{0}=s_{0}(\lambda_{1}+\lambda_{2}+1)$. The series terms $j_{3}=[\Delta(s_{0}:j_{2})]$ and $j_{4}=[\Delta(s_{0}:j_{1})]$ are denoted comprising of $s_{0}$ and $\delta_{0}$ terms, respectively, where the series expression $[\Delta(\sigma:\mathcal{L}_{z})]$ with $z$ terms is defined as
\begin{align}
\label{a6}
[\Delta(\sigma:\mathcal{L}_{z})]=\Delta(\sigma:\mathcal{L}_{1}), \Delta(\sigma:\mathcal{L}_{2}), \cdots , \Delta(\sigma:\mathcal{L}_{z}).
\end{align}

\subsection{PDF and CDF of SNR for the Eavesdropper Channels}

\subsubsection{Eavesdropper at the RF link}
Similar to the main RF channel, the PDF of SNR for $S-E_{1}$ link can be defined as \cite[Eq.~(3),]{dacosta2007average}
\begin{align}
\label{a7}
f_{\gamma_{r_{e}}}(\gamma)=\mathcal{C}\sum_{N_{e}=1}^{2}\sum_{w=0}^{\mu_{e}-1}X_{N_{e},w}\gamma^{\mu_{e}-w-1}e^{-l_{N_{e}}\gamma},
\end{align}
where $\mathcal{C}=\frac{k_{e}^{\mu_{e}}}{K_{e}^{\mu_{e}}\Gamma(\mu_{e})}$, $X_{1,w}=\frac{\Gamma(\mu_{e}+w)\mu_{e}^{\mu_{e}-w}}{w!\Gamma(\mu_{e}-w)4^{w}\phi_{s,e}^{\mu_{e}-w}K_{e}^{w}}(-1)^{w}$, $X_{2,w}=\frac{\Gamma(\mu_{e}+w)\mu_{e}^{\mu_{e}-w}}{w!\Gamma(\mu_{e}-w)4^{w}\phi_{s,e}^{\mu_{e}-w}K_{e}^{w}}(-1)^{\mu_{e}}$, $l_{1}=\frac{2\mu_{e}(k_{e}-K_{e})}{\phi_{s,e}}$, and $l_{2}=\frac{2\mu_{e}(k_{e}+K_{e})}{\phi_{s,e}}$. Considering the specific condition of $0<\eta_{e}<\infty$, parameters $k_{e}$ and $K_{e}$ are denoted as $k_{e}=\frac{2+\eta_{e}^{-1}+\eta_{e}}{4}$ and $K_{e}=\frac{\eta_{e}^{-1}-\eta_{e}}{4}$. Parameter $\mu_{e}>0$ denotes the fading of $S-E_{1}$ channel. Similar to \eqref{a3}, CDF of $\gamma_{r_{e}}$ is expressed as \cite[Eq.~(4),]{yang2018physical}
\begin{align}  
\label{a8}
F_{\gamma_{r_{e}}}(\gamma)=1-\mathcal{C}\sum_{N_{e}=1}^{2}\sum_{w=0}^{\mu_{e}-1}\sum_{y=0}^{\mu_{e}-w-1}\frac{\gamma^{y}}{y!}e^{-l_{N_{e}}\gamma}l_{N_{e}}^{y}Y_{N_{e},w},
\end{align}
where $Y_{1,w}=\frac{\Gamma(\mu_{e}+w)(-1)^{w}K_{e}^{-w}}{w!2^{\mu_{e}+w}(k_{e}-K_{e})^{\mu_{e}-w}}$, and $Y_{2,w}=\frac{\Gamma(\mu_{e}+w)(-1)^{\mu_{e}}K_{e}^{-w}}{w!2^{\mu_{e}+w}(k_{e}+K_{e})^{\mu_{e}-w}}$.
\vspace{2mm}
\subsubsection{Eavesdropper at the FSO link}
The $R-E_{2}$ link experiences DGG fading similar to main FSO link, the PDF of $\gamma_{d_{e}}$ is defined as \cite[Eq.~(12),]{alquwaiee2015performance}
\begin{align}
\label{a9}
f_{\gamma_{d_{e}}}(\gamma)=\frac{\mathcal{B}_{1}}{s_{e}\gamma}G_{1,\lambda_{1}+\lambda_{2}+1}^{\lambda_{1}+\lambda_{2}+1,0}\biggl[\mathcal{B}_{2}t^{\tau}\biggl(\frac{\gamma}{U_{e}}\biggl)^{\frac{\tau}{s_{e}}}\biggl |
 \begin{array}{c}
 j_{2}\\
 j_{1}\\
\end{array}\biggl],
\end{align}
where $s_{e}$ represents the two detection types through which $E_{2}$ receives the optical signals ($s_{e}=1$ denotes HD technique and $s_{e}=2$ denotes IM/DD technique). As in \eqref{a4}, all parameters of DGG fading model related to atmospheric turbulence and pointing error are similar for both FSO main and eavesdropper channels. The electrical SNR of this link is addressed as $U_{e}=\frac{\left\{\Lambda_{e}E[I_{e}]\right\}^{s_{e}}}{P_{e}}$. Here, $\Lambda_{e}$, $I_{e}$, and $P_{e}$ denote photoelectric conversion coefficient, receiver irradiance, and number of sample apertures, respectively, for $R-E_{2}$ link. As a result, $\frac{E[I_{e}^{2}]}{E[I_{e}]^{2}}\triangleq\iota_{e}^{2}+1$ indicates the trivial relationship between $\phi_{r,e}$ and $U_{e}$, where $\iota_{e}^{2}$ is the scintillation index for $R-E_{2}$ link. Similar to \eqref{a5}, CDF of $\gamma_{d_{e}}$ is defined as \cite[Eq.~(14),]{alquwaiee2015performance}
\begin{align}
\label{a10}
F_{\gamma_{d_{e}}}(\gamma)=\mathcal{B}_{5}G_{s_{e}+1,\delta_{e}+1}^{\delta_{e},1}\biggl[\mathcal{B}_{6}\biggl(\frac{\gamma}{U_{e}}\biggl)^{\tau}\biggl |
 \begin{array}{c}
 1, j_{5}\\
 j_{6}, 0\\
\end{array}\biggl],
\end{align}
where $\mathcal{B}_{5}=\frac{\epsilon^{2}\lambda_{2}^{b_{1}-\frac{1}{2}}\lambda_{1}^{b_{2}-\frac{1}{2}}(2\pi)^{1-\frac{s_{e}(\lambda_{1}+\lambda_{2})}{2}}s_{e}^{b_{1}+b_{2}-2}}{a_{2}\lambda_{1}\Gamma(b_{1})\Gamma(b_{2})}$, $B_{6}=\left(\frac{\mathcal{B}_{2}t^{a_{2}\lambda_{1}}}{s_{e}^{\lambda_{1}+\lambda_{2}}}\right)^{s_{e}}$, and $\delta_{e}=s_{e}(\lambda_{1}+\lambda_{2}+1)$. The series terms $j_{5}=[\Delta(s_{e}:j_{2})]$ and $j_{6}=[\Delta(s_{e}:j_{1})]$ comprise of $s_{e}$ and $\delta_{e}$ terms, respectively, and are symbolized according to \eqref{a6}.

\subsection{PDF and CDF of SNR for Dual-hop RF-FSO Link}
Utilizing order statistics, the CDF of $\gamma_{d}$ is expressed as \cite[Eq.~(5),]{zedini2014performance}
\begin{align}
\nonumber
F_{\gamma_{d}}(\gamma)&=\Pr\left[\min(\gamma_{r_{0}}, \gamma _{d_{0}})<\gamma\right]
\\
\label{a11}
&=F_{\gamma_{r_{0}}}(\gamma)+F_{\gamma_{d_{0}}}(\gamma)-F_{\gamma_{r_{0}}}(\gamma)F_{\gamma_{d_{0}}}(\gamma).
\end{align}
Placing \eqref{a3} and \eqref{a5} into \eqref{a11} and performing some arithmatic simplifications, the CDF of $\gamma_{d}$ is written as
\begin{align}
\label{a12}
F_{\gamma_{d}}(\gamma)=1-\mathcal{A}\sum_{N_{0}=1}^{2}\sum_{v=0}^{\mu_{0}-1}\sum_{x=0}^{\mu_{0}-v-1}\frac{\gamma^{x}}{x!}e^{-l_{N_{0}}\gamma}l_{N_{0}}^xY_{N_{0},v}\left\{1-\mathcal{B}_{3}G_{s_{0}+1,\delta_{0}+1}^{\delta_{0},1}\biggl[\mathcal{B}_{4}\biggl(\frac{\gamma}{U_{d}}\biggl)^{\tau}\biggl |
 \begin{array}{c}
 1, j_{3}\\
 j_{4}, 0\\
\end{array}\biggl]\right\}.
\end{align}
The PDF of $\gamma_{d}$ is defined as
\cite[Eq.~(4),]{gupta2018performance}
\begin{align}
\label{a13}
f_{\gamma_{d}}(\gamma)=f_{\gamma_{r_{0}}}(\gamma)+f_{\gamma_{d_{0}}}(\gamma)-f_{\gamma_{r_{0}}}(\gamma)F_{\gamma_{d_{0}}}(\gamma)-f_{\gamma_{d_{0}}}(\gamma)F_{\gamma_{r_{0}}}(\gamma).
\end{align}
Substituting \eqref{a2}, \eqref{a3}, \eqref{a4}, and \eqref{a5} into \eqref{a13} and doing some simplifications, the PDF of $\gamma_{d}$ is obtained as
\begin{align}
\nonumber
f_{\gamma_{d}}(\gamma)=&\mathcal{A}\sum_{N_{0}=1}^{2}\sum_{v=0}^{\mu_{0}-1}e^{-l_{N_{0}}\gamma}\left\{\frac{\mathcal{B}_{1}}{s_{0}}  \sum_{x=0}^{\mu_{0}-v-1}\frac{l_{N_{0}}^x}{x!}\gamma^{x-1}Y_{N_{0},v}G_{1,\lambda_{1}+\lambda_{2}+1}^{\lambda_{1}+\lambda_{2}+1,0}\left[\mathcal{B}_{2}t^{\tau}\left(\frac{\gamma}{U_{d}}\right)^\frac{\tau}{s_{0}]}\biggl |
\begin{array}{c}
j_{2} \\
j_{1} \\
\end{array}
\right]\right.
\\
\label{a14}
&+\left.\mathcal{B}_{3} X_{N_{0},v} G_{s_{0}+1,\delta_{0}+1}^{s_{0}+1,0}\biggl[\mathcal{B}_{4}\biggl(\frac{\gamma}{U_{d}}\biggl)^{\tau}\biggl |
\begin{array}{c}
j_{3},1 \\
0,j_{4} \\
\end{array}
\biggl]\right\}.
\end{align}

\section{Performance Analysis}
In this section, we derive novel closed-form expressions of SOP and SPSC for both the proposed scenarios. We also derive asymptotic expressions for SOP to obtain better intuition on our analysis.

\subsection{Secure Outage Probability}
SOP is a decisive performance metric for secrecy analysis. It demonstrates the reverse mechanism to evaluate secrecy performance. For both \textit{scenario-1} and \textit{scenario-2} of our proposed model, we derive two different SOP expressions based on different positions of the eavesdropper.

\subsubsection{Scenario-1}
Considering $\mathcal{T}_{C_{1}}$ as the target secrecy rate for \textit{scenario-1}, the occurrence of an outage event for secrecy measurement can be defined when $\mathcal{T}_{D_{1}}$ falls below $\mathcal{T}_{C_{1}}$. According to this theory, SOP for RF-FSO combined system in \textit{scenario-1} can be introduced as \cite[Eq.~(14),]{lei2015performance}
\begin{align}
\nonumber
SOP_{1}&=\Pr\left\{\mathcal{T}_{D_{1}}\leq \mathcal{T}_{C_{1}}\right\}
\\
\label{a15}
&=\Pr\left\{\gamma_{d}\leq\varphi_{1}\gamma_{r_{e}}+\varphi_{1}-1\right\}.
\end{align}
where $\varphi_{1}=2^{\mathcal{T}_{C_{1}}}$. The term in \eqref{a15} can be described as \cite{sumona2021security, badrudduza2021security}
\begin{align}
\nonumber
SOP_{1}&=1-\int_{0}^{\infty}\int_{\varphi_{1}\gamma_{r_{e}}+\varphi_{1}-1}^{\infty} f_{d}(\gamma_{d})f_{r_{e}}(\gamma_{r_{e}})d\gamma_{d}d\gamma_{r_{e}}
\\
\label{a16}
&=\int_{0}^{\infty}F_{\gamma_{d}}(\varphi_{1}\gamma+\varphi_{1}-1)f_{\gamma_{r_{e}}}(\gamma)d\gamma,
\end{align}
Although the expression defined in \eqref{a16} is the exact expression of SOP, solving \eqref{a16} in closed-form is not possible due to mathematical complexities. Hence, for DF relaying scheme, we often consider lower-bound SOP as \cite[Eq.~(6),]{lei2015physical}
\begin{align}
\label{a17}
SOP_{1}\geq SOP_{1,L}=\int_{0}^{\infty}F_{\gamma_{d}}(\varphi_{1}\gamma)f_{\gamma_{r_{e}}}(\gamma)d\gamma.
\end{align}
Substituting \eqref{a7} and \eqref{a12} into \eqref{a17}, the SOP for \textit{scenario-1} is expressed as
\begin{align}
\label{a18}
SOP_{1,L}=&1-\mathcal{A}\mathcal{C}\sum_{N_{0}=1}^{2}\sum_{N_{e}=1}^{2}\sum_{v=0}^{\mu_{0}-1}\sum_{w=0}^{\mu_{e}-1}\sum_{x=0}^{\mu_{0}-v-1}\frac{l_{N_{0}}^x}{x!}X_{N_{e},w}Y_{N_{0},v}(Q_{1}-\mathcal{B}_{3}Q_{2}).
\end{align}
Performing integration operation utilizing \cite[Eq.~(3.351.3),]{gradshteyn2014table}, the term $Q_{1}$ in \eqref{a18} is obtained as
\begin{align}
\nonumber
Q_{1}&=\int_{0}^{\infty}\gamma_{r_{e}}^{z_{1}-1}e^{-\mathcal{F}\gamma_{r_{e}}}\varphi_{1}^{x}d\gamma_{r_{e}}
\\
&=\frac{\varphi_{1}^{x}\Gamma(z_{1})}{(\mathcal{F})^{z_{1}}},
\end{align}
where $\mathcal{F}=\varphi_{1}l_{N_{0}}+l_{N_{e}}$, and $z_{1}=\mu_{e}-w+x$. Subsequently, converting the exponential term to Meijer's $G$ function and then performing integration via applying \cite[eqs. (2.24.1.1) and (8.4.3.1),]{Prudnikov1992integrals}, the term $Q_{2}$ in \eqref{a18} is obtained as
\begin{align}
\nonumber
Q_{2}=&\int_{0}^{\infty}\gamma_{r_{e}}^{z_{1}-1}\varphi_{1}^{x}e^{-\mathcal{F}\gamma_{r_{e}}}G_{s_{0}+1,\delta_{0}+1}^{\delta_{0},1}\left[\mathcal{B}_{4}\biggl(\frac{\varphi_{1}\gamma_{r_{e}}}{U_{d}}\biggl)^{\tau}\biggl |
\begin{array}{c}
 1,j_{3} \\
 j_{4},0 \\
\end{array}
\right]d\gamma_{r_{e}}
\\
\nonumber
=&\int_{0}^{\infty}\gamma_{r_{e}}^{z_{1}-1}\varphi_{1}^{x}G_{0,1}^{1,0}\biggl[\mathcal{F}\gamma_{r_{e}}\biggl |
 \begin{array}{c}
 -\\
 0\\
\end{array}\biggl]G_{s_{0}+1,\delta_{0}+1}^{\delta_{0},1}\left[\mathcal{B}_{4}\biggl(\frac{\varphi_{1}\gamma_{r_{e}}}{U_{d}}\biggl)^{\tau}\biggl |
\begin{array}{c}
 1,j_{3} \\
 j_{4},0 \\
\end{array}
\right]d\gamma_{r_{e}}
\\
=&\frac{\varphi_{1}^{x}}{\mathcal{F}^{z_{1}}}G_{s_{0}+2,\delta_{0}+1}^{\delta_{0},2}\left[\mathcal{B}_{4}\biggl(\frac{\varphi_{1}\tau}{U_{d}\mathcal{F}}\biggl)^{\tau}\biggl |
\begin{array}{c}
 1,1-z_{1},j_{3} \\
 j_{4},0 \\
\end{array}
\right].
\end{align}
The expression in \eqref{a18} can be utilized to obtain Rayleigh-$\Gamma\Gamma$ distribution \cite[Eq.~(15),]{abd2017physical} by setting $\eta_{0}=\eta_{e}=1$, $\mu_{0}=\mu_{e}=1$, $a_{1}=a_{2}=\Omega_{1}=\Omega_{2}=1$. It can also be reformed as (Nakagami-$m$)-$\Gamma\Gamma$ distribution \cite[Eq.~(13),]{lei2017secrecy} via setting $\eta_{0}\geq1$, $\eta_{e}\geq1$, $\mu_{0}\geq1$, $\mu_{e}\geq1$, and $a_{1}=a_{2}=\Omega_{1}=\Omega_{2}=1$.

\quad 

\noindent
\textbf{Asymptotic Expression:}

To gain better understanding of the secrecy incident of our proposed model, we derive asymptotic SOP expression by setting $U_{d}\rightarrow\infty$. By converting the Meijer's $G$ term described in \eqref{a18} via utilizing \cite[Eq.~(6.2.2),]{springer1979algebra} and \cite[Eq.~(19),]{ansari2015performance}, the asymptotic SOP for \textit{scenario-1} is expressed as
\begin{align}
\nonumber
SOP_{1,\infty}=1-&\mathcal{A}\mathcal{C}\sum_{N_{0}=1}^{2}\sum_{N_{e}=1}^{2}\sum_{v=0}^{\mu_{0}-1}\sum_{w=0}^{\mu_{e}-1}\sum_{x=0}^{\mu_{0}-v-1}\frac{l_{N_{0}}^x}{x!}X_{N_{e},w}Y_{N_{0},v}
\\
\label{b2}
&\times\left[Q_{1}-\frac{\mathcal{B}_{3}\varphi_{1}^{x}}{\mathcal{F}^{z_{1}}}\sum_{p=1}^{\delta_{0}}\Gamma (j_{4,p})\mathcal{B}_{4}^{j_{4,p}}\frac{\prod_{h=1,h\neq p}^{\delta_{0}} \Gamma(j_{4,h}-j_{4,p})}{\prod_{h=3}^{s_{0}+2}\Gamma(j_{3,h}-j_{4,p})} \biggl(\frac{U_{d}\mathcal{F}}{\varphi_{1}\tau}\biggl)^{-\tau j_{4,p}}\right].
\end{align}
\vspace{1mm}
\subsubsection{scenario-2}
For the DF based relaying configuration setup in \textit{scenario-2} of Fig. \ref{f1}, SOP is defined as
\begin{align}
\label{a19}
SOP_{2}=\Pr\left\{ \mathcal{T}_{D_{2}} < \mathcal{T}_{C_{2}} \right\},
\end{align}
where $T_{C_{2}}$ is defined as the target SC for second scenario of our proposed model. As the model explained in \textit{scenario-2} is autonomous and links RF-FSO relaying system, SOP for this case can be defined by substituting \eqref{b1} into \eqref{a19} as
\begin{align} 
\nonumber
SOP_{2}&=\Pr\left\{min(\mathcal{T}_{S},\mathcal{T}_{R})<T_{C_{2}}\right\}
\\
\nonumber
&=1-\Pr\left\{min(\mathcal{T}_{S},\mathcal{T}_{R})\geq T_{C_{2}}\right\}
\\
\label{a20}
&=1-\Pr\left\{\mathcal{T}_{S}\geq T_{C_{2}}\right\}\Pr\left\{\mathcal{T}_{R}\geq T_{C_{2}}\right\}.
\end{align}
Substituting the values of \eqref{a3}, \eqref{a5}, and \eqref{a9} into \eqref{a20}, we have
\begin{align}
\label{a21}
SOP_{2}=\int_{0}^{\infty}F_{\gamma_{d_{0}}}(\varphi_{2}\gamma+\varphi_{2}-1)f_{\gamma_{d_{e}}}(\gamma)\left\{1-F_{\gamma_{r_{0}}}(\varphi_{2}-1)\right\}d\gamma+F_{\gamma_{r_{0}}}(\varphi_{2}-1),
\end{align}
where $\varphi_{2}=2^{2T_{C_{2}}}$. For defining closed-form expression, we must delineate the lower bound of SOP, similar to \eqref{a17}, as
\begin{align}
\label{a22}
SOP_{2}\geq SOP_{2,L}\cong\int_{0}^{\infty}F_{\gamma_{d_{0}}}(\varphi_{2}\gamma)f_{\gamma_{d_{e}}}(\gamma)\left\{1-F_{\gamma_{r_{0}}}(\varphi_{2}-1)\right\}d\gamma+F_{\gamma_{r_{0}}}(\varphi_{2}-1).
\end{align}
Placing \eqref{a3}, \eqref{a5}, and \eqref{a9} into \eqref{a22} and performing some integration and simplifications via utilizing {\cite[Eq.~(2.24.1.1),]{Prudnikov1992integrals}}, \eqref{a22} is obtained as
\begin{align}
\nonumber
SOP_{2,L}=1-&\mathcal{A}\sum_{N_{0}=1}^{2}\sum_{v=0}^{\mu_{0}-1}\sum_{x=0}^{\mu_{0}-v-1}\frac{(\varphi_{2}-1)^{x}}{x!}e^{-l_{N_{0}}(\varphi_{2}-1)}l_{N_{0}}^xY_{N_{0},v}
\\
\label{a23}
&\times\left\{1-\mathcal{B}_{3}\mathcal{B}_{5}G_{s_{e}+\delta_{0}+1,s_{0}+\delta_{e}+1}^{\delta_{e}+1,\delta_{0}}\left[\frac{\mathcal{B}_{6}}{\mathcal{B}_{4}}\biggl(\frac{U_{d}}{U_{e}\varphi_{2}}\biggl)^{\tau}\biggl|
\begin{array}{c}
 1-j_{4},1,j_{5} \\
 j_{6},0,1-j_{3} \\
\end{array}
\right]\right\}.
\end{align}
It is observed that the derived expression in \eqref{a23} can be utilized to generate Rayleigh-$\Gamma\Gamma$ distribution \cite[Eq.~(19),]{pan2019secrecy} while considering the conditions $\eta_{0}=1$, and $\mu_{0}=1$, $a_{1}=a_{2}=\Omega_{1}=\Omega_{2}=1$.

\quad

\noindent
\textbf{Asymptotic Expression:}

Similar to \eqref{b2}, we define the asymptote for \textit{scenario-2} in our proposed model to improve our analysis via extracting better intuitions. Making use of \cite[Eq.~(29),]{pattanayak2020physical}, asymptotic expression of \eqref{a23} is derived as
\begin{align}
\nonumber
SOP_{2,\infty}=1-&\mathcal{A}\sum_{N_{0}=1}^{2}\sum_{v=0}^{\mu_{0}-1}\sum_{x=0}^{\mu_{0}-v-1}\frac{(\varphi_{2}-1)^{x}}{x!}e^{-l_{N_{0}}(\varphi_{2}-1)}l_{N_{0}}^{x}Y_{N_{0},v}\left\{1-\mathcal{B}_{1}\mathcal{B}_{5}\right.
\\
\label{a24}
&\times\left.\sum_{p=1}^{\delta_{0}}\frac{\prod_{h=1,h\neq p}^{\delta_{0}}\Gamma(\mathcal{J}_{1,p}-\mathcal{J}_{1,h})\prod_{h=1}^{\delta_{e}+1}\Gamma(1+\mathcal{J}_{2,h}-\mathcal{J}_{1,p})}{\prod_{h=\delta_{0}+1}^{s_{e}+\delta_{0}+1}\Gamma(1+\mathcal{J}_{1,h}-\mathcal{J}_{1,p})\prod_{h=\delta_{e}+2}^{s_{0}+\delta_{e}+1}\Gamma(\mathcal{J}_{1,p}-\mathcal{J}_{2,h})}\left[\frac{\mathcal{B}_{6}}{\mathcal{B}_{4}}\biggl(\frac{U_{d}}{U_{e}\varphi_{2}}\biggl)^{\tau}\right]^{\mathcal{J}_{1,p}-1}\right\},
\end{align}
where $\mathcal{J}_{1}=(1-j_{4},1,j_{5})$, and $\mathcal{J}_{2}=(j_{6},0,1-j_{3})$.

\subsection{Strictly Positive Secrecy Capacity}
The probability of SPSC is the inverse probable term of outage probability that is a positive volume of secrecy capacity. In this subsection, we derive expressions of SPSC for both scenarios described in Fig. \ref{f1}.

\subsubsection{scenario-1}
Considering \textit{scenario-1} wherein eavesdropper $E_{1}$ is located near the $S-R$ link, SPSC can be expressed as \cite{badrudduza2020enhancing, ibrahim2021enhancing}
\begin{align}
\nonumber
SPSC_{1}&=\Pr(\mathcal{T}_{D_{1}}>0)
\\
\nonumber
&=\Pr(\gamma_{d}>\gamma_{r_{e}})
\\
\nonumber
&=\int_0^{\infty}\int_0^{\gamma_{d}}f_{d}(\gamma_{d})f_{r_{e}}(\gamma_{r_{e}})d\gamma_{r_{e}}d\gamma_{d}
\\
\label{a25}
&=\int_{0}^{\infty}f_{\gamma_{d}}(\gamma)F_{\gamma_{r_{e}}}(\gamma)d\gamma.
\end{align}
Substituting \eqref{a8} and \eqref{a14} into \eqref{a25} and employing mathematical simplifications, SPSC for \textit{scenario-1} is obtained as
\begin{align}
\nonumber
SPSC_{1}=&\mathcal{A}\sum_{N_{0}=1}^{2}\sum_{v=0}^{\mu_{0}-1}\left\{X_{N_{0},v}\mathcal{R}_{1}+\sum_{x=0}^{\mu_{0}-v-1}\frac{l_{N_{0}}^x}{x!}Y_{N_{0},v}\mathcal{R}_{2}-\mathcal{C}\sum_{N_{e}=1}^{2}\sum_{w=0}^{\mu_{e}-1}\sum_{y=0}^{\mu_{e}-w-1}\frac{l_{N_{e}}^y}{y!}Y_{N_{e},w}\right.
\\
\label{a26}
&\times\left.\left[X_{N_{0},v}\mathcal{R}_{3}+\sum_{x=0}^{\mu_{0}-v-1}\frac{l_{N_{0}}^x}{x!}Y_{N_{0},v}\mathcal{R}_{4}\right]\right\},
\end{align}
where $\mathcal{R}_{1}$, $\mathcal{R}_{2}$, $\mathcal{R}_{3}$, and $\mathcal{R}_{4}$ are four integration terms. Now, Utilizing {\cite[eqs.~(2.24.1.1) and (8.4.3.1),]{Prudnikov1992integrals}} $\mathcal{R}_{1}$ is obtained as
\begin{align}
\nonumber
\mathcal{R}_{1}=&\int_{0}^{\infty}\mathcal{B}_{3}\gamma_{d}^{z_{2}-1}e^{-l_{N_{0}}\gamma_{d}}G_{s_{0}+1,\delta_{0}+1}^{\delta_{0}+1,0}\left[\mathcal{B}_{4}\biggl(\frac{\gamma_{d}}{U_{d}}\biggl)^{\tau}\biggl |
\begin{array}{c}
 j_{3},1 \\
 0,j_{4} \\
\end{array}
\right] d\gamma_{d}
\\
\nonumber
=&\int_{0}^{\infty}\mathcal{B}_{3}\gamma_{d}^{z_{2}-1}G_{0,1}^{1,0}\left[l_{N_{0}}\gamma_{d}\biggl |
\begin{array}{c}
 - \\
 0 \\
\end{array}
\right]G_{s_{0}+1,\delta_{0}+1}^{\delta_{0}+1,0}\left[\mathcal{B}_{4}\biggl(\frac{\gamma_{d}}{U_{d}}\biggl)^{\tau}\biggl |
\begin{array}{c}
 j_{3},1 \\
 0,j_{4} \\
\end{array}
\right] d\gamma_{d}
\\
=&\frac{\mathcal{B}_{3}}{l_{N_{0}}^{z_{2}}}G_{s_{0}+2,\delta_{0}+1}^{\delta_{0}+1,1}\left[\mathcal{B}_{4}\biggl(\frac{\tau}{U_{d}l_{N_{0}}}\biggl)^{\tau}\biggl |
\begin{array}{c}
 1-z_{2},j_{3},1 \\
 0,j_{4} \\
\end{array}
\right],
\end{align}
where $z_{2}=\mu_{0}-v$. The second integration term $\mathcal{R}_{2}$ is calculated similarly as
\begin{align}
\nonumber
\mathcal{R}_{2}=&\int_{0}^{\infty}\frac{\mathcal{B}_{1}}{s_{0}}\gamma_{d}^{x-1}e^{-l_{N_{0}}\gamma_{d}}G_{1,\lambda_{1}+\lambda_{2}+1}^{\lambda_{1}+\lambda_{2}+1,0}\left[\mathcal{B}_{2}t^{\tau}\biggl(\frac{\gamma_{d}}{U_{d}}\biggl)^{\frac{\tau}{s_{0}}}\biggl |
\begin{array}{c}
 j_{2} \\
 j_{1} \\
\end{array}
\right] d\gamma_{d}
\\
\nonumber
=&\int_{0}^{\infty}\mathcal{B}_{1}s_{0}^{-1}\gamma_{d}^{x-1}G_{0,1}^{1,0}\left[l_{N_{0}}\gamma_{d}\biggl |
\begin{array}{c}
 - \\
 0 \\
\end{array}
\right]G_{1,\lambda_{1}+\lambda_{2}+1}^{\lambda_{1}+\lambda_{2}+1,0}\left[\mathcal{B}_{2}t^{\tau}\biggl(\frac{\gamma_{d}}{U_{d}}\biggl)^{\frac{\tau}{s_{0}}}\biggl |
\begin{array}{c}
 j_{2} \\
 j_{1} \\
\end{array}
\right] d\gamma_{d}
\\
=&\frac{\mathcal{B}_{3}}{l_{N_{0}}^x}G_{s_{0}+1,\delta_{0}}^{\delta_{0},1}\left[\mathcal{B}_{4}\biggl(\frac{\tau}{U_{d}l_{N_{0}}}\biggl)^{\tau}\biggl |
\begin{array}{c}
 1-x,j_{3} \\
 j_{4} \\
\end{array}
\right].
\end{align}
Performing identical mathematical calculations as $\mathcal{R}_{1}$ and $\mathcal{R}_{2}$, $\mathcal{R}_{3}$ yields to
\begin{align}
\nonumber
\mathcal{R}_{3}=&\int_{0}^{\infty}\mathcal{B}_{3}\gamma_{d}^{z_{3}-1}e^{-\mathcal{H}\gamma_{d}}G_{s_{0}+1,\delta_{0}+1}^{\delta_{0}+1,0}\left[\mathcal{B}_{4}\biggl(\frac{\gamma_{d}}{U_{d}}\biggl)^{\tau}\biggl |
\begin{array}{c}
 j_{3},1 \\
 0,j_{4} \\
\end{array}
\right] d\gamma_{d}
\\
\nonumber
=&\int_{0}^{\infty}\mathcal{B}_{3}\gamma_{d}^{z_{3}-1}G_{0,1}^{1,0}\left[\mathcal{H}\gamma_{d}\biggl |
\begin{array}{c}
 - \\
 0 \\
\end{array}
\right]G_{s_{0}+1,\delta_{0}+1}^{\delta_{0}+1,0}\left[\mathcal{B}_{4}\biggl(\frac{\gamma_{d}}{U_{d}}\biggl)^{\tau}\biggl |
\begin{array}{c}
 j_{3},1 \\
 0,j_{4} \\
\end{array}
\right] d\gamma_{d}
\\
=&\frac{\mathcal{B}_{3}}{\mathcal{H}^{z_{3}}}G_{s_{0}+2,\delta_{0}+1}^{\delta_{0}+1,1}\left[\mathcal{B}_{4}\biggl(\frac{\tau}{U_{d}\mathcal{H}}\biggl)^{\tau}\biggl |
\begin{array}{c}
 1-z_{3},j_{3},1 \\
 0,j_{4} \\
\end{array}
\right],
\end{align}
where $\mathcal{H}=l_{N_{0}}+l_{N_{e}}$, and $z_{3}=z_{2}+y$. The last integration term $\mathcal{R}_{4}$ is obtained similarly as
\begin{align}
\nonumber
\mathcal{R}_{4}=&\int_{0}^{\infty}\frac{\mathcal{B}_{1}}{s_{0}}\gamma_{d}^{z_{4}-1}e^{-\mathcal{H}\gamma_{d}}G_{1,\lambda_{1}+\lambda_{2}+1}^{\lambda_{1}+\lambda_{2}+1,0}\left[\mathcal{B}_{2}t^{\tau}\biggl(\frac{\gamma_{d}}{U_{d}}\biggl)^{\frac{\tau}{s_{0}}}\biggl |
\begin{array}{c}
 j_{2} \\
 j_{1} \\
\end{array}
\right] d\gamma_{d}
\\
\nonumber
=&\int_{0}^{\infty}\mathcal{B}_{1}s_{0}^{-1}\gamma_{d}^{z_{4}-1}G_{0,1}^{1,0}\left[\mathcal{H}\gamma_{d}\biggl |
\begin{array}{c}
 - \\
 0 \\
\end{array}
\right]G_{1,\lambda_{1}+\lambda_{2}+1}^{\lambda_{1}+\lambda_{2}+1,0}\left[\mathcal{B}_{2}t^{\tau}\biggl(\frac{\gamma_{d}}{U_{d}}\biggl)^{\frac{\tau}{s_{0}}}\biggl |
\begin{array}{c}
 j_{2} \\
 j_{1} \\
\end{array}
\right] d\gamma_{d}
\\
=&\frac{\mathcal{B}_{3}}{\mathcal{H}^{z_{4}}}G_{s_{0}+1,\delta_{0}}^{\delta_{0},1}\left[\mathcal{B}_{4}\biggl(\frac{\tau}{U_{d}\mathcal{H}}\biggl)^{\tau}\biggl |
\begin{array}{c}
 1-z_{4},j_{3} \\
 j_{4} \\
\end{array}
\right],
\end{align}
where $z_{4}=x+y$.

\subsubsection{scenario-2}
For the DF based system in \textit{scenario-2}, SPSC can be defined as \cite[Eq.~(9),]{liu2012probability}
\begin{align}
\nonumber
SPSC_{2}&=\Pr\left[min(\mathcal{T}_{S},\mathcal{T}_{R})>0\right]
\\
\label{a27}
&=\Pr\left(\mathcal{T}_{S}>0\right)\Pr\left(\mathcal{T}_{R}>0\right).
\end{align}
For $S-R$ link, in the case of \textit{scenario-2}, the positive probability term is expressed as
\begin{align}
\nonumber
\Pr\left(\mathcal{T}_{S}>0\right)&=\Pr\left[\frac{1}{2}\log_{2}(1+\gamma_{r_{0}})>0\right]
\\
\nonumber
&=\Pr\left(\gamma_{r_{0}}>0\right)
\\
\label{a28}
&=1.
\end{align}
For second-hop, the probability term is defined as
\begin{align}
\nonumber
\Pr\left(\mathcal{T}_{R}>0\right)&=\Pr\left\{\frac{1}{2}\left[\log_{2}(1+\gamma_{d_{0}})-\log_{2}(1+\gamma_{d_{e}})\right]>0\right\}
\\
\nonumber
&=\Pr\left(\gamma_{d_{0}}>\gamma_{d_{e}}\right)
\\
\label{a29}
&=1-\int_{0}^{\infty}F_{\gamma_{d_{0}}}(\gamma)f_{\gamma_{d_{e}}}(\gamma)d\gamma.
\end{align}
Substituting \eqref{a28} and \eqref{a29} into \eqref{a27}, the SPSC can be denoted as
\begin{align}
\label{a30}
SPSC_{2}=1-\int_{0}^{\infty}F_{\gamma_{d_{0}}}(\gamma)f_{\gamma_{d_{e}}}(\gamma)d\gamma.
\end{align}
Setting the values of \eqref{a5} and \eqref{a9} into \eqref{a30}, and performing integration utilizing {\cite[Eq.~(2.24.1.1),]{Prudnikov1992integrals}}, \eqref{a30} is obtained as
\begin{align}
\label{a31}
SPSC_{2}&=1-\mathcal{B}_{3}\mathcal{B}_{5}G_{s_{e}+\delta_{0}+1,s_{0}+\delta_{e}+1}^{\delta_{e}+1,\delta_{0}}\left[\frac{\mathcal{B}_{6}}{\mathcal{B}_{4}}\biggl(\frac{U_{d}}{U_{e}}\biggl)^{\tau}\biggl|
\begin{array}{c}
 1-j_{4},1,j_{5} \\
 j_{6},0,1-j_{3} \\
\end{array}
\right].
\end{align}
It can be noted that by setting the fading parameter values to $\eta_{0}=1, \mu_{0}=1$ and $a_{1}=a_{2}=\Omega_{1}=\Omega_{2}=1$, the expression derived in \eqref{a31} matches with \cite[Eq.~(23),]{pan2019secrecy} of the Rayleigh-$\Gamma\Gamma$ fading distribution.

\section{Numerical Results}

\begin{figure}[!b]
\vspace{-25mm}
    \centerline{\includegraphics[width=0.6\textwidth,angle=0]{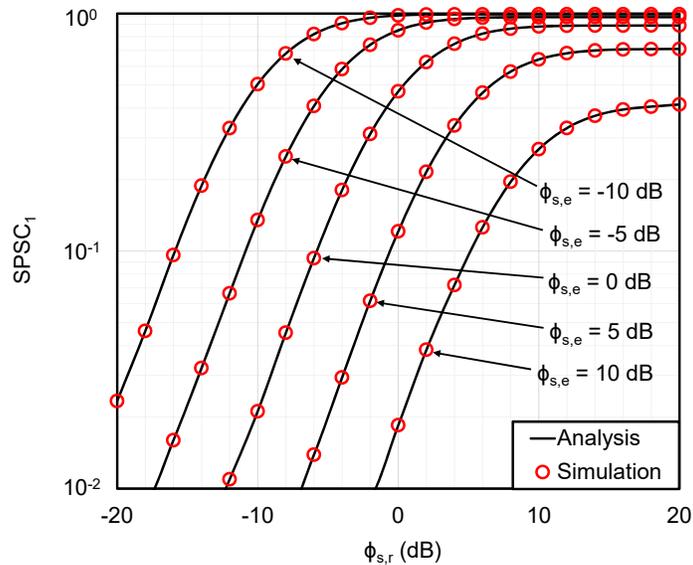}}
    \vspace{-30mm }
    \caption{
         The SPSC$_{1}$ versus $\phi_{s,r}$ for selected values of $\phi_{s,e}$ with $\eta_{0}=\eta_{e}=20$, $\mu_{0}=\mu_{e}=2$, $a_{1}=a_{2}=2.1$, $b_{1}=4$, $b_{2}=4.5$, $\Omega_{1}=1.07$, $\Omega_{2}=1.06$, $\lambda_{1}=\lambda_{2}=1$, $s_{0}=1$, $U_{d}=10$ dB, and $\epsilon=1$.
    }
    \label{g1}
\end{figure}

In this section, we present some analytical results utilizing the derived expressions of the secrecy metrics, namely lower bound and asymptotic SOP, and SPSC to demonstrate the impact of the system parameters on secrecy performance considering both eavesdropping scenarios. To corroborate our analytical outcomes, we also demonstrate Monte-Carlo simulations via generating $\eta-\mu$ and DGG random variables in MATLAB and averaging 100,000 channel realizations to acquire each value of the secrecy parameters. It is noteworthy in figures that the analytical and simulation results are in good agreement with each other. The analysis is performed by assuming some parametric values such as $\eta_{0}\geq0$, $\eta_{e}\geq0$, $\mu_{0}\geq0$, $\mu_{e}\geq0$, $\mathcal{T}_{D_{1}}=\mathcal{T}_{D_{2}}=1$, $\mathcal{T}_{C_{1}}=\mathcal{T}_{C_{2}}=0.5$ bits/sec/Hz, $s_{0}=s_{0}=(1,2)$, and $\epsilon=\{1, 6.7\}$. To analyze natural turbulence levels over the DGG link, we set up the following values for atmospheric turbulence parameters \cite{kashani2015novel}.
\begin{itemize}
\item $a_{1}=1.86$, $a_{2}=1$, $b_{1}=0.5$, $b_{2}=1.8$, $\Omega_{1}=1.51$, $\Omega_{2}=1$, $\lambda_{1}=17$, and $\lambda_{2}=9$ for strong turbulence (ST).
\item $a_{1}=2.17$, $a_{2}=1$, $b_{1}=0.55$, $b_{2}=2.35$, $\Omega_{1}=1.58$, $\Omega_{2}=0.97$, $\lambda_{1}=28$, and $\lambda_{2}=13$ for moderate turbulence (MT).
\item $a_{1}=a_{2}=2.1$, $b_{1}=4$, $b_{2}=4.5$, $\Omega_{1}=1.07$, $\Omega_{2}=1.06$, and $\lambda_{1}=\lambda_{2}=1$ for weak turbulence (WT).
\end{itemize}

\begin{figure}[!ht]
\vspace{-25mm}
    \centerline{\includegraphics[width=0.6\textwidth,angle=0]{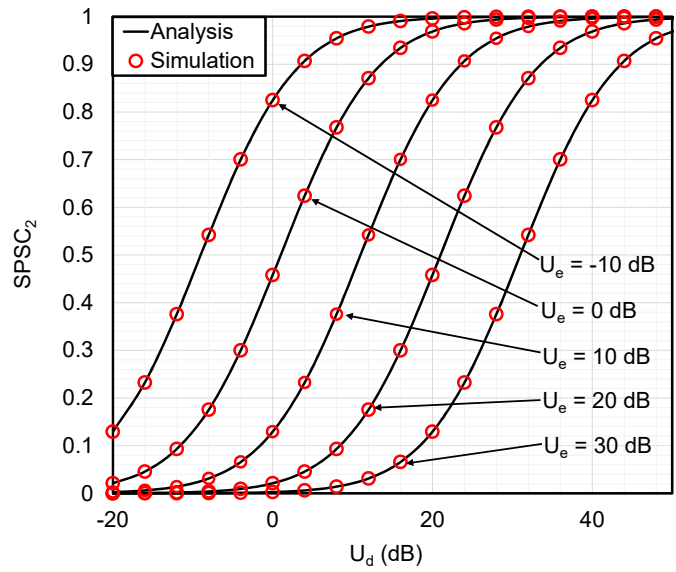}}
        \vspace{-30mm}
    \caption{
         The SPSC$_{2}$ versus $U_{d}$ for selected values of $U_{e}$ with $a_{1}=1.86$, $a_{2}=1$, $b_{1}=0.5$, $b_{2}=1.8$, $\Omega_{1}=1.51$, $\Omega_{2}=1$, $\lambda_{1}=17$, $\lambda_{2}=9$, $s_{0}=s_{e}=1$, and $\epsilon=1$.
    }
    \label{g2}
\end{figure}

The impact of average SNR of the eavesdropper links on the secrecy performance is investigated in Figs. \ref{g1} and \ref{g2}.

Fig. \ref{g1} indicates relationship between SPSC$_{1}$ and $\phi_{s,r}$ i.e. the first scenario. It is observed SPSC$_{1}$ increases when the average SNR of $S-E_{1}$ link $\phi_{s,e}$ decreases from $10$ dB to $-10$ dB. On the other hand, Fig. \ref{g2} illustrates the effect of average SNR of $R-E_{2}$ link, i.e., the second scenario. This time $U_{e}$ is decreased from $30$ dB to $-10$ dB. As a result, the performance metric SPSC$_{2}$ that is plotted against $U_{d}$, increases remarkably. These two events reveal that decrease in $\phi_{s,e}$ and $U_{e}$ renders the eavesdropper channels weaker relative to the main channel thereby the SPSC performance improves as reported in \cite{sarker2020secrecy, juel2021secrecy}.

\begin{figure}[!ht]
\vspace{-25mm}
    \centerline{\includegraphics[width=0.6\textwidth,angle=0]{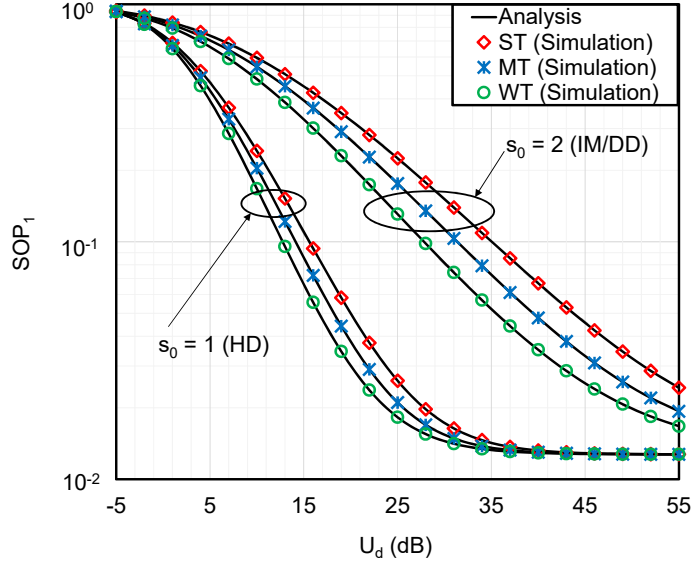}}
        \vspace{-30mm}
    \caption{
         The SOP$_{1}$ versus $U_{d}$ for selected values of $s_{0}$, $a_{1}$, $a_{2}$, $b_{1}$, $b_{2}$, $\Omega_{1}$, $\Omega_{2}$, $\lambda_{1}$, and $\lambda_{2}$ with $\eta_{0}=\eta_{e}=50$, $\mu_{0}=\mu_{e}=3$, $\phi_{s,r}=10$ dB, $\phi_{s,e}=0$ dB, and $\epsilon=1$.
    }
    \label{g3}
\end{figure}

\begin{figure}[!ht]
\vspace{-25mm}
    \centerline{\includegraphics[width=0.6\textwidth,angle=0]{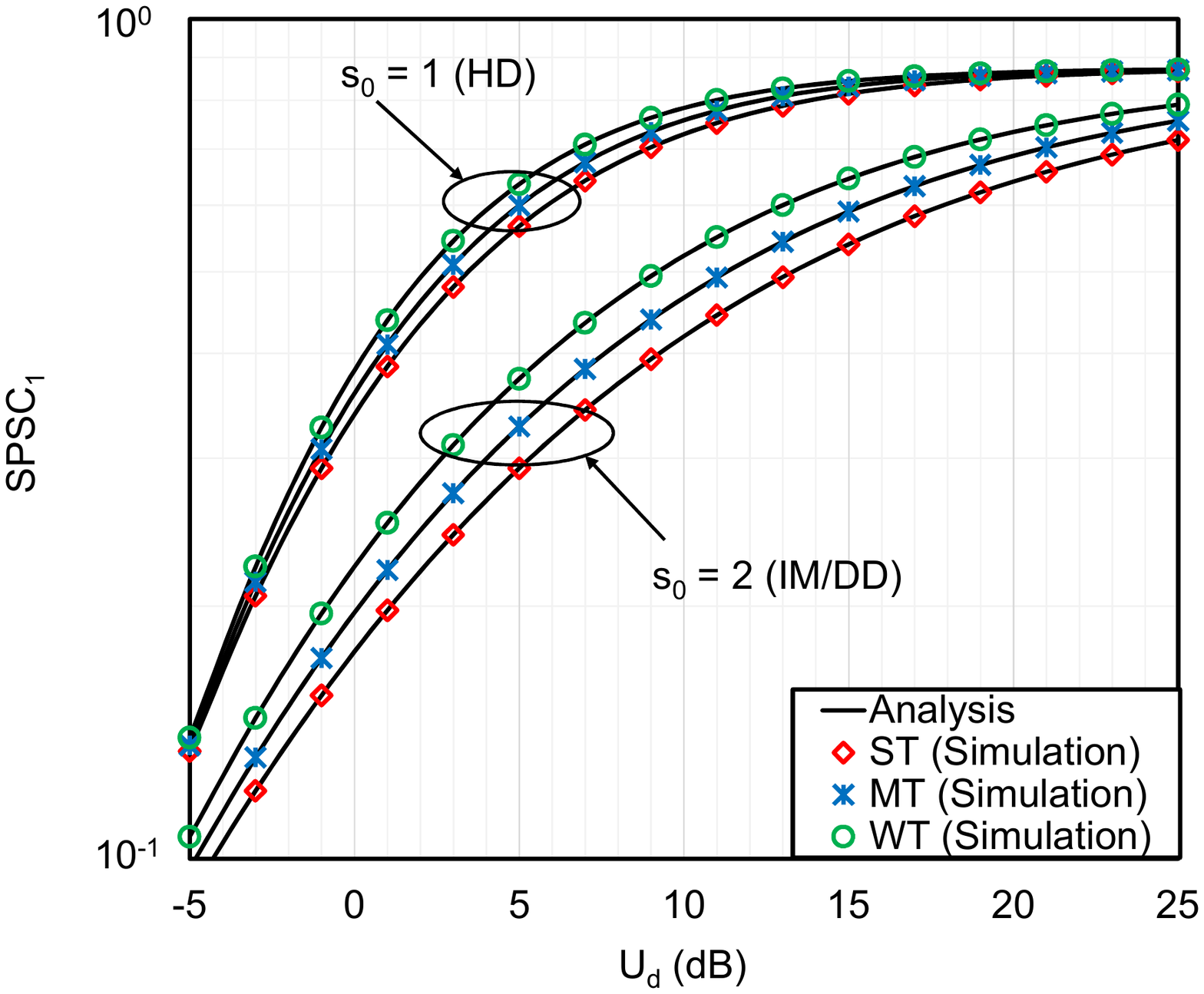}}
        \vspace{-30mm}
    \caption{
         The SPSC$_{1}$ versus $U_{d}$ for selected values of $s_{0}$, $a_{1}$, $a_{2}$, $b_{1}$, $b_{2}$, $\Omega_{1}$, $\Omega_{2}$, $\lambda_{1}$, and $\lambda_{2}$ with $\eta_{0}=\eta_{e}=25$, $\mu_{0}=\mu_{e}=2$, $\phi_{s,r}=5$ dB, $\phi_{s,e}=0$ dB, and $\epsilon=1$.
    }
    \label{g4}
\end{figure}

\begin{figure}[!ht]
\vspace{-25mm}
    \centerline{\includegraphics[width=0.6\textwidth,angle=0]{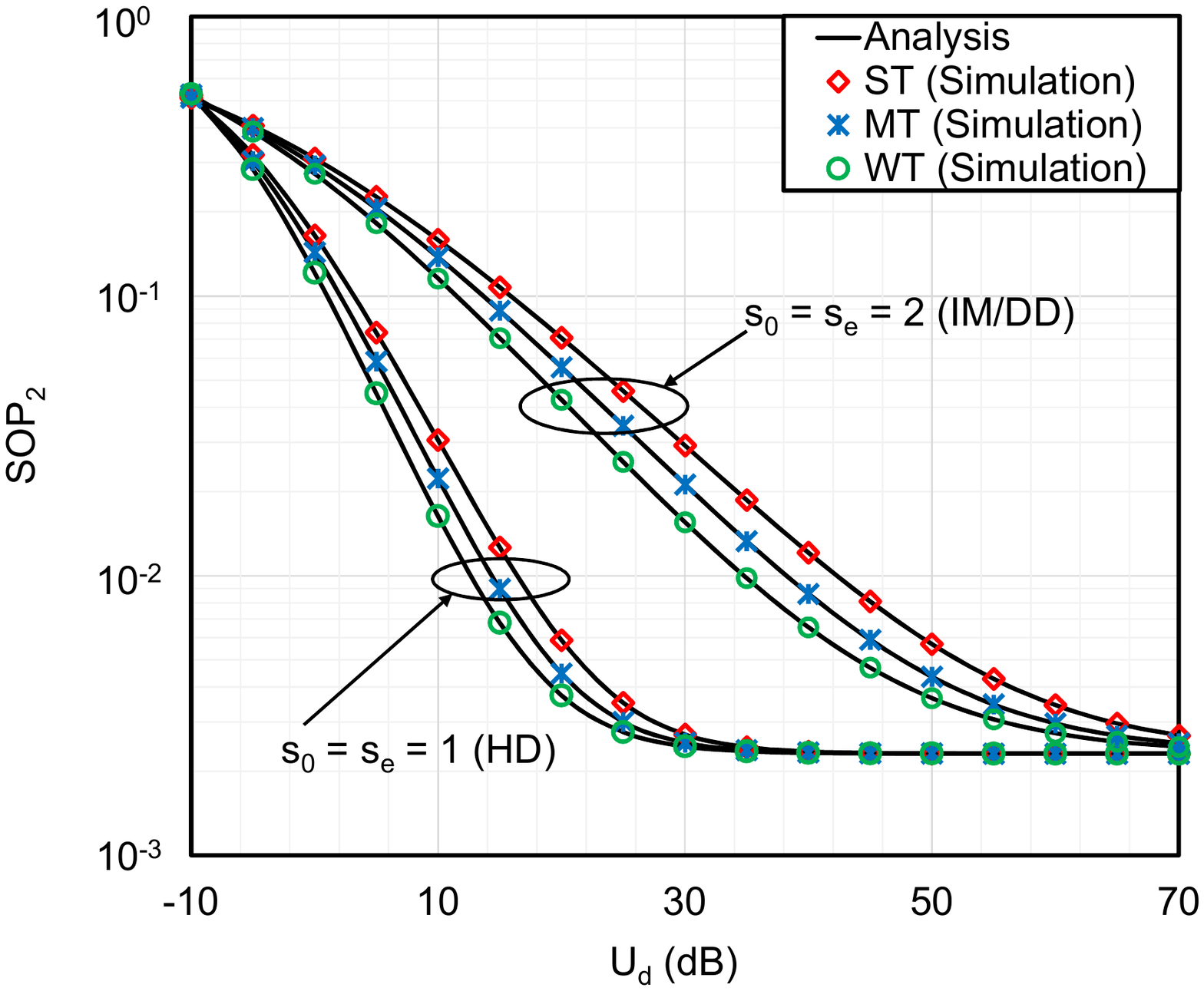}}
        \vspace{-30mm}
    \caption{
         The SOP$_{2}$ versus $U_{d}$ for selected values of $s_{0}$, $s_{e}$, $a_{1}$, $a_{2}$, $b_{1}$, $b_{2}$, $\Omega_{1}$, $\Omega_{2}$, $\lambda_{1}$, and $\lambda_{2}$ with $\eta_{0}=5$, $\mu_{0}=1$, $\phi_{s,r}=12$ dB, $U_{e}=-10$ dB, and $\epsilon=1$.
    }
    \label{g5}
\end{figure}

A comparative analysis between two types of detection techniques at the receiver is demonstrated in Figs. \ref{g3}-\ref{g5} where all the figures are plotted against $U_{d}$.
Figs. \ref{g3} and \ref{g4} demonstrate this comparison under \textit{Scenario-1} while Fig. \ref{g5} illustrates the same under \textit{Scenario-2}. Results imply that utilizing HD technique ($s_{0}=s_{e}=1$) for signal detection provides better secrecy output relative to IM/DD technique ($s_{0}=s_{e}=2$). The reason behind these outcomes is that HD technique provides higher SNR than IM/DD technique at the destination receiver. The results demonstrated in \cite{lei2018secrecy, pan2019secrecy} also agree with our demonstrated results that validates our analysis.

\begin{figure}[!ht]
\vspace{-25mm}
    \centerline{\includegraphics[width=0.6\textwidth,angle=0]{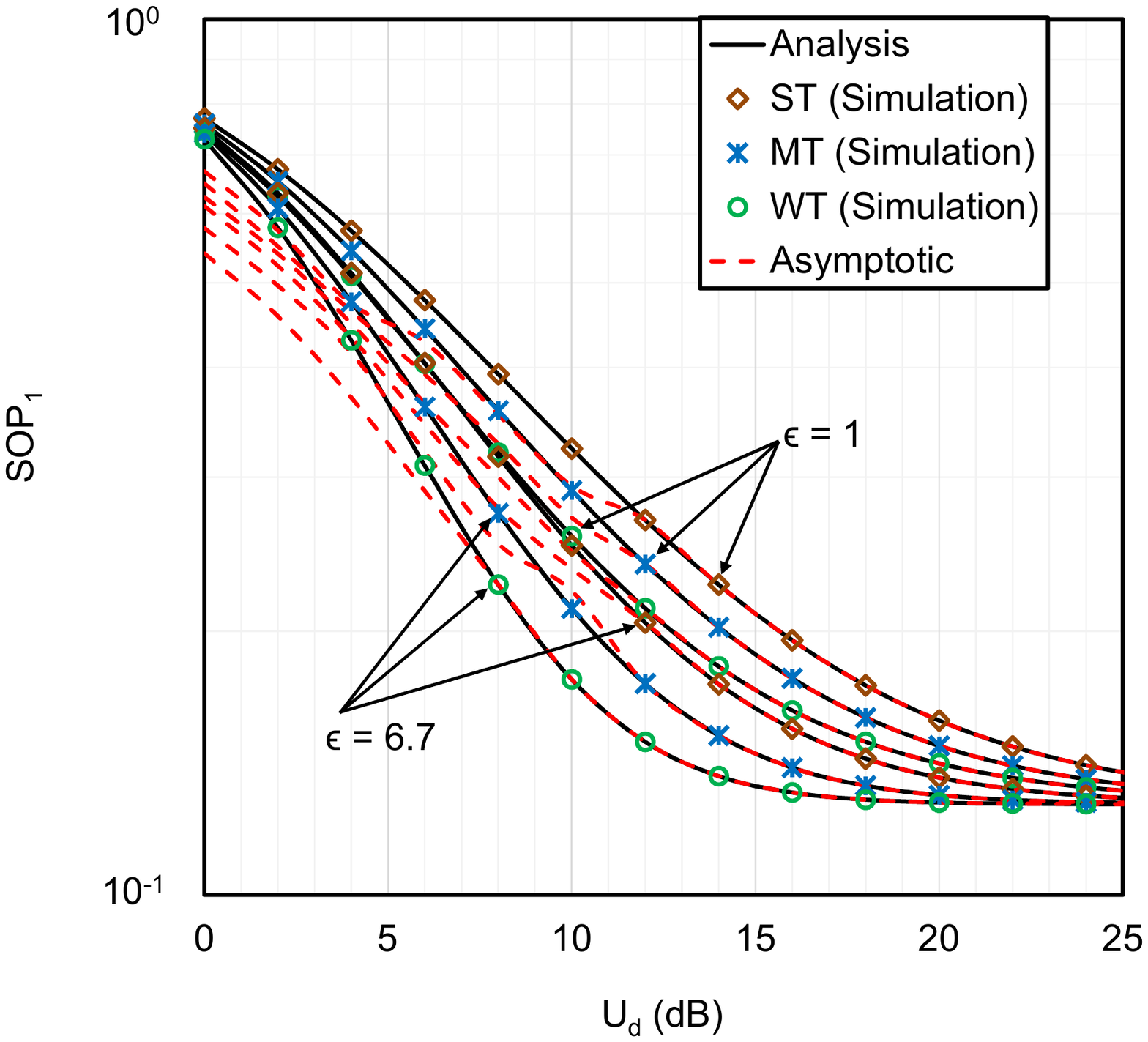}}
        \vspace{-30mm}
    \caption{
         The SOP$_{1}$ versus $U_{d}$ for selected values of $\epsilon$, $a_{1}$, $a_{2}$, $b_{1}$, $b_{2}$, $\Omega_{1}$, $\Omega_{2}$, $\lambda_{1}$, and $\lambda_{2}$ with $\eta_{0}=\eta_{e}=25$, $\mu_{0}=\mu_{e}=4$, $\phi_{s,r}=5$ dB, $\phi_{s,e}=0$ dB, and $s_{0}=1$.
    }
    \label{g6}
\end{figure}

\begin{figure}[!ht]
\vspace{-25mm}
    \centerline{\includegraphics[width=0.6\textwidth,angle=0]{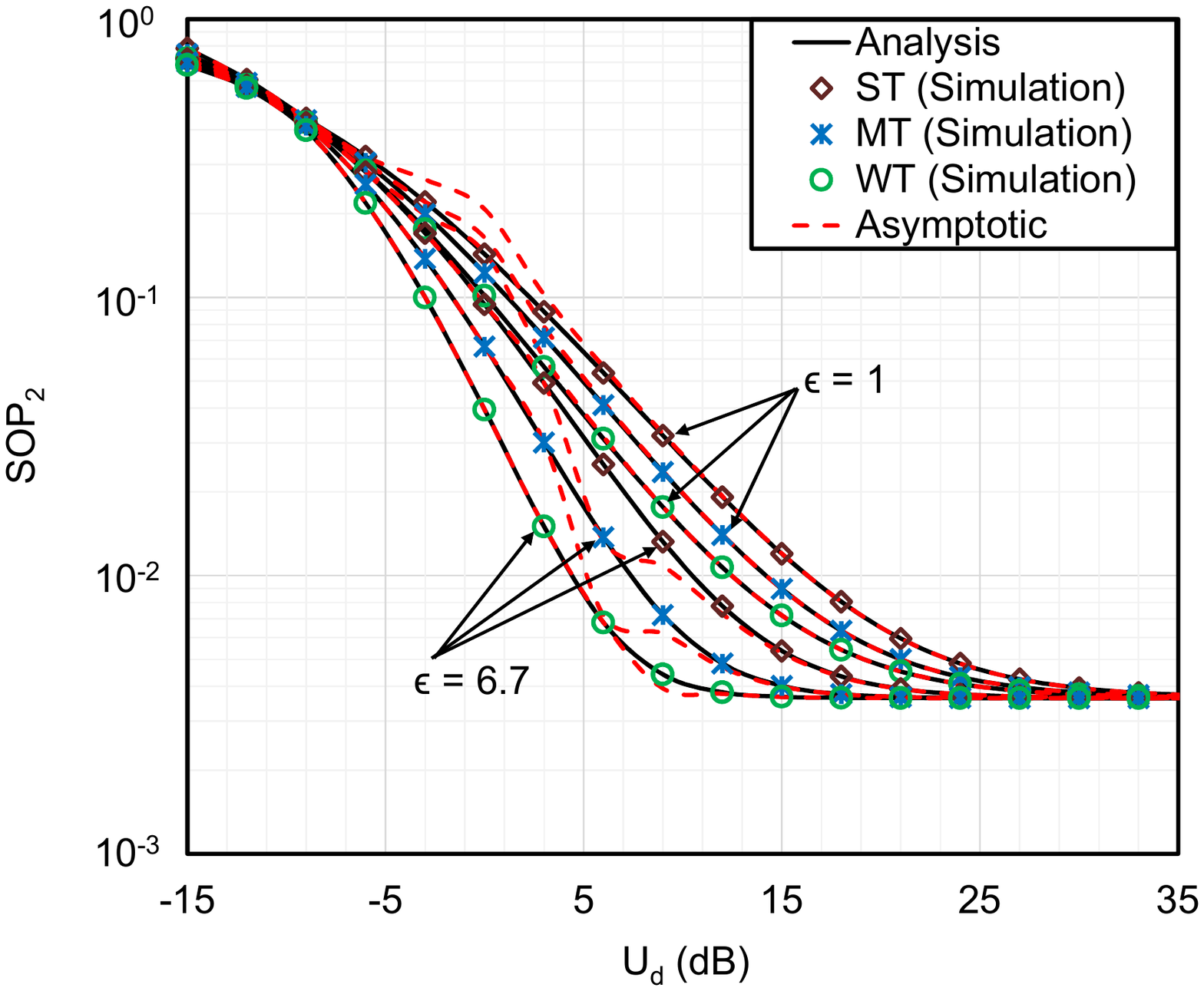}}
        \vspace{-30mm}
    \caption{
         The SOP$_{2}$ versus $U_{d}$ for selected values of $\epsilon$, $a_{1}$, $a_{2}$, $b_{1}$, $b_{2}$, $\Omega_{1}$, $\Omega_{2}$, $\lambda_{1}$, and $\lambda_{2}$ with $\eta_{0}=2$, $\mu_{0}=1$, $\phi_{s,r}=10$ dB, $s_{0}=s_{e}=1$, and $U_{e}=-12$ dB.
    }
    \label{g7}
\end{figure}

\begin{figure}[!ht]
\vspace{-25mm}
    \centerline{\includegraphics[width=0.6\textwidth,angle=0]{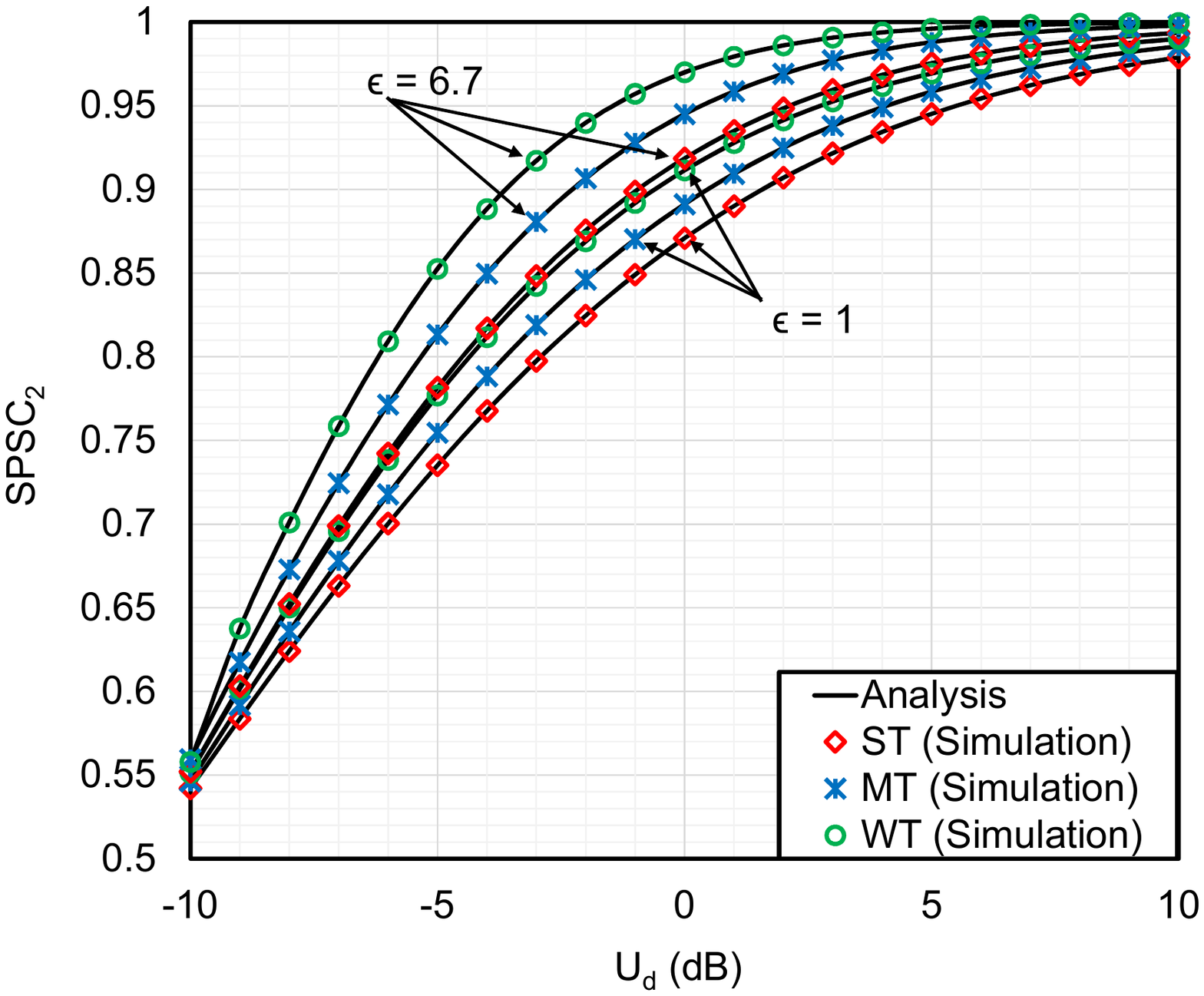}}
        \vspace{-30mm}
    \caption{
         The SPSC$_{2}$ versus $U_{d}$ for selected values of $\epsilon$, $a_{1}$, $a_{2}$, $b_{1}$, $b_{2}$, $\Omega_{1}$, $\Omega_{2}$, $\lambda_{1}$, and $\lambda_{2}$ with $s_{0}=s_{e}=1$ and $U_{e}=-10$ dB.
    }
    \label{g8}
\end{figure}

The influence of pointing errors in FSO links is demonstrated in Figs. \ref{g6}-\ref{g8}.
Fig. \ref{g6} illustrates SOP$_{1}$ vs $\phi_{s,r}$ applicable to \textit{scenario-1}, and Figs. \ref{g7} and \ref{g8} demonstrates SOP$_{2}$ and SPSC$_{2}$, respectfully, plotted against $U_{d}$ applicable to \textit{scenario-2}. All three figures demonstrate that the secrecy performance for both eavesdropping scenarios increases when DGG main link undergoes from severe pointing error ($\epsilon$=1) to negligible pointing error ($\epsilon$=6.7). Similar outcomes were demonstrated in \cite{lei2017secrecy, pan2019secrecy} that corroborate our investigations. Besides, asymptotic outputs are demonstrated in Figs. \ref{g6} and \ref{g7} for SOP$_{1}$ and SOP$_{2}$, respectively, that reveal the asymptotic lower bound SOP results can tightly approximate our derived lower bound SOP results in high SNR regime.

Besides the detection technique types and pointing errors, the turbulence parameters of DGG channel also place notable influences on the secrecy performance. Figs. \ref{g3}-\ref{g8} indicate the effects of three turbulence conditions, namely, ST, MT, and WT. Our demonstrated outcomes show the expected results similar to \cite{lei2017secrecy, lei2018secrecy, pan2019secrecy} that secrecy performance with weaker turbulence in Figs. \ref{g3}-\ref{g8} clearly outperforms that under stronger turbulence.

\quad

\noindent
\textbf{Generalization Demonstrated via the Proposed Model:}

\begin{figure}[!ht]
\vspace{-25mm}
    \centerline{\includegraphics[width=0.6\textwidth,angle=0]{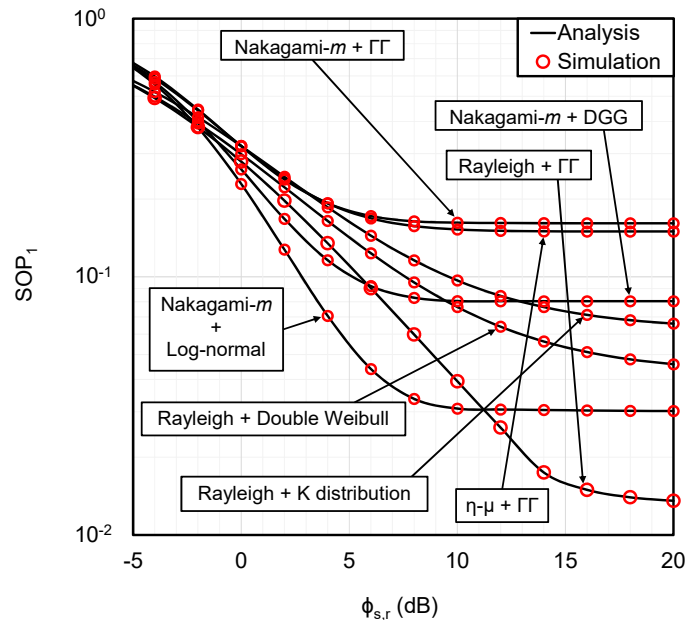}}
    \vspace{-30mm}
    \caption{
         The SOP$_{1}$ versus $\phi_{s,r}$ for selected values of $\eta_{0}$, $\eta_{e}$, $\mu_{0}$, $\mu_{e}$, $a_{1}$, $a_{2}$, $b_{1}$, $b_{2}$, $\Omega_{1}$, $\Omega_{2}$, $\lambda_{1}$, and $\lambda_{2}$ with $\phi_{s,e}=-5$ dB, $s_{0}=1$, $U_{d}=5$ dB, and $\epsilon=6.7$.
    }
    \label{g9}
\end{figure}

\begin{figure}[!ht]
\vspace{-25mm}
    \centerline{\includegraphics[width=0.6\textwidth,angle=0]{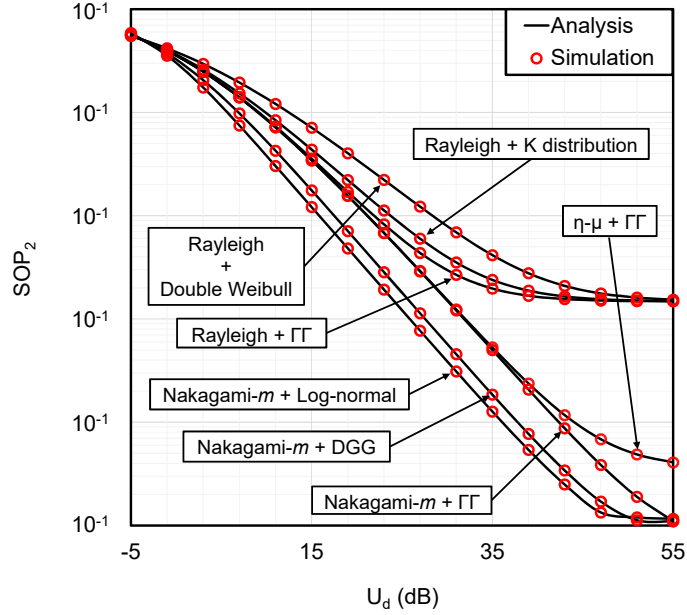}}
        \vspace{-30mm}
    \caption{
         The SOP$_{2}$ versus $U_{d}$ for selected values of $\eta_{0}$, $\mu_{0}$, $a_{1}$, $a_{2}$, $b_{1}$, $b_{2}$, $\Omega_{1}$, $\Omega_{2}$, $\lambda_{1}$, and $\lambda_{2}$ with $\phi_{s,r}=12$ dB, $s_{0}=s_{e}=1$, $U_{e}=-5$ dB, and $\epsilon=1$.
    }
    \label{g10}
\end{figure}

In this proposed model, we assume $\eta-\mu$ fading model over the RF hop and DGG turbulence model over the FSO hop. The $\eta-\mu$ distribution has an outstanding generic nature that can emerge several multipath fading channels as special cases that are indicated in Table \ref{t1}.
\begin{table*}[!ht]
\centering
\caption{Some researches as Special Cases of Our Proposed Model}
\scalebox{0.90}{%
\begin{tabular}{>{\centering\arraybackslash}p{1.3 cm}|>{\centering\arraybackslash} p{7cm}| >{\centering\arraybackslash}p{8cm} |>{\centering\arraybackslash}p{1.8cm} }
\hline
Reference & RF link & FSO link & Eavesdropper link
\\ 
\hline
\hline
- & Rayleigh ($\eta=1$, $\mu=1$) & $K$ distribution ($a_{1}=a_{2}=b_{1}=\Omega_{1}=\Omega_{2}=\lambda_{1}=\lambda_{2}=1, b_{2}=1.8$) & -
\\
\hline
- & Rayleigh ($\eta=1$, $\mu=1$) & Double Weibull ($a_{1}=a_{2}=2.1, b_{1}=b_{2}=1$, $\Omega_{1}=1.07, \Omega_{2}=1.06$, $\lambda_{1}=\lambda_{2}=1$)) & -
\\
\hline
- & Nakagami-$m$ ($\eta=20$, $\mu=2$) & Log-normal ($a_{1}=a_{2}=0.01$, $b_{1}=4$, $b_{2}=4.5$, $\Omega_{1}=\Omega_{2}=1.07$, $\lambda_{1}=\lambda_{2}=1$) & -
\\
\hline
\cite{pattanayak2018statistical} & Nakagami-$m$ ($\eta=20$, $\mu=2$) & DGG ($a_{1}=2.17$, $a_{2}=1, b_{1}=0.55$, $b_{2}=2.35$, $\Omega_{1}=1.58$, $\Omega_{2}=0.97$, $\lambda_{1}=28$, $\lambda_{2}=13$) & -
\\
\hline
\cite{sharma2016decode} & $\eta-\mu$ ($\eta=100$, $\mu=2$) & $\Gamma\Gamma$ ($a_{1}=a_{2}=\Omega_{1}=\Omega_{2}=\lambda_{1}=\lambda_{2}=1$, $b_{1}=2.296$, $b_{2}=1.822$) & -
\\
\hline 
\cite{lei2017secrecy} & Nakagami-$m$ ($\eta=20$, $\mu=2$) & $\Gamma\Gamma$ ($a_{1}=a_{2}=\Omega_{1}=\Omega_{2}=\lambda_{1}=\lambda_{2}=1$, $b_{1}=2.296$, $b_{2}=1.822$) & RF
\\
\hline
\cite{abd2017physical} & Rayleigh ($\eta=1$, $\mu=1$) & $\Gamma\Gamma$ ($a_{1}=a_{2}=\Omega_{1}=\Omega_{2}=\lambda_{1}=\lambda_{2}=1$, $b_{1}=2.296$, $b_{2}=1.822$) & RF
\\
\hline 
\cite{pan2019secrecy} & Rayleigh ($\eta=1$, $\mu=1$) & $ \Gamma\Gamma$ ($a_{1}=a_{2}=\Omega_{1}=\Omega_{2}=\lambda_{1}=\lambda_{2}=1$, $b_{1}=2.296$, $b_{2}=1.822$) & FSO
\\
\hline
\end{tabular}}
\label{t3}
\end{table*}
On the other hand, DGG turbulence model is also regarded as a generalized FSO turbulent model from which multiple classical FSO models can be generated as special cases, as listed in Table \ref{t2}. It can be clearly observed the demonstrated channel models indicated in \cite{lei2017secrecy} and \cite{abd2017physical} can be addressed as the special cases of \textit{scenario-1}. Likewise, the secure models denoted in \cite{pan2019secrecy} can be addressed as a special case of our proposed \textit{scenario-2}. Subsequently, Table \ref{t3} summarizes some other special cases that are not available in the literature till date and these are graphically represented in Figs. \ref{g9} and \ref{g10}.

Hence, we can express that existing researches in \cite{lei2017secrecy, abd2017physical, pan2019secrecy} can be obtained as the special cases of our model that clearly demonstrates the novelty and supremacy of our model relative to the existing literature. 

\section{Conclusion}
This paper analyses the secrecy performance of an RF-FSO mixed framework under eavesdropping attempts via the RF or FSO links. The RF hop experiences $\eta-\mu$ fading channel whereas the FSO hop undergoes unified DGG turbulence with pointing error impairments. The secrecy analyses are performed deducing expressions for SPSC and SOP in closed-form and obtaining further useful insights via deriving asymptotic SOP expressions. All the analytical expressions are also verified via MC simulations. Utilizing the derived expressions, impacts of fading, weak to strong atmospheric turbulences, and pointing errors are also observed. It is seen that our demonstrated results exhibit a generalization of the various reported outcomes in the literature. Moreover, a comparison between HD and IM/DD techniques reveals the HD technique offers a better and secure outage performance over the proposed scheme relative to the IM/DD technique.
\bibliography{main.bib}

\end{document}